\documentclass[a4paper,fleqn]{cas-dc}

\usepackage[numbers]{natbib}
\usepackage{multirow,slashed,hyperref,cleveref}
\hypersetup{colorlinks=true,linkcolor=blue,citecolor=magenta,filecolor=magenta, urlcolor=cyan}
\def\ie{{\it i.e.}}

\newcommand{\be}{\begin{equation}}
\newcommand{\ee}{\end{equation}}
\def\bsp#1\esp{\begin{split}#1\end{split}}

\def\lag{{\cal L}}
\def\sss{\scriptscriptstyle}

\def\uR{u_R}
\def\uRb{\bar{u}_R}

\newcommand{\fr}{{\sc FeynRules}}
\newcommand{\ma}{{\sc MadAnalysis}~5}
\newcommand{\maddm}{{\sc MadDM}}

\newcommand{\mg}{{\sc MG5\_aMC}}
\newcommand{\micromegas}{{\sc Mi\-crO\-ME\-GAs}}

\def\dmsimpt{{\tt DMSimpt}}

\begin{document}
\def\floatpagepagefraction{1}
\def\textpagefraction{.001}
\shorttitle{$t$-channel dark matter models}

\title [mode = title]{Closing in on $t$-channel simplified dark matter models}                      

\author[1]{Chiara Arina}
\author[2,3]{Benjamin Fuks}
\author[1]{Luca Mantani}
\author[4]{Hanna Mies}
\author[5,6]{Luca Panizzi}
\author[5]{Jakub Salko}
\shortauthors{C.~Arina {\it et~al.}}
\shorttitle{{Closing in on $t$-channel simplified dark matter models}}

\address[1]{Centre for Cosmology, Particle Physics and Phenomenology (CP3), Universit\'e catholique de Louvain, B-1348 Louvain-la-Neuve, Belgium}
\address[2]{Sorbonne Universit\'e, CNRS, Laboratoire de Physique Th\'eorique et  Hautes \'Energies, LPTHE, F-75005 Paris, France}
\address[3]{Institut Universitaire de France, 103 boulevard Saint-Michel, F-75005 Paris, France}
\address[4]{Institute for Theoretical Particle Physics and Cosmology, RWTH Aachen University, D-52056 Aachen, Germany}
\address[5]{Department of Physics and Astronomy, Uppsala University, Box 516, SE-751 20 Uppsala, Sweden}
\address[6]{School of Physics and Astronomy, University of Southampton, Highfield, Southampton SO17 1BJ, UK}

\begin{abstract}
A comprehensive analysis of cosmological and collider constraints is presented for three simplified models characterised by a dark matter candidate (real scalar, Majorana fermion and real vector) and a coloured mediator (fermion, scalar and fermion respectively) interacting with the right-handed up quark of the Standard Model. Constraints from dark matter direct and indirect detection and relic density are combined with bounds originating from the re-interpretation of a full LHC run~2 ATLAS search targeting final states with multiple jets and missing transverse energy. Projections for the high-luminosity phase of the LHC are also provided to assess future exclusion and discovery reaches, which show that analogous future search strategies will not allow for a significant improvement compared with the present status.
From the cosmological point of view, we demonstrate that thermal dark matter is largely probed (and disfavoured) by constraints from current direct and indirect detection experiments. These bounds and their future projections have moreover the potential of probing the whole parameter space when combined with the expectation of the high-luminosity phase of the LHC.
\end{abstract}


\maketitle

\section{Introduction}
The nature of dark matter and the way it is connected to the Standard Model (SM)
is one of the most puzzling issues in particle physics today. Dark matter
searches consequently hold a central place in the present astroparticle and
particle physics program. However, despite of convincing indirect evidence for
its existence~\cite{Bertone:2010zza}, dark matter still evades any direct
detection probes. Experimental searches at colliders, in underground nuclear
recoil experiments and with gamma-ray telescopes therefore put stronger and
stronger constraints on the
viability of any dark matter model. Those bounds are very often explored, in a
model-independent approach, as limits on a set of simplified models for
dark matter phenomenology. In those simplified models, the dark matter is
considered as a massive particle whose interactions with the SM
arise through a mediator particle. In so-called $s$-channel setups~\cite{
Fox:2012ru,Haisch:2013ata,Backovic:2015soa}, the mediator is
a colour singlet and couples to a pair of either dark matter or SM particles. On
the contrary, in a $t$-channel configuration, the mediator interacts instead
with one SM state and the dark matter~\cite{Arina:2020udz}.

In this work, we consider three simplified $t$-channel scenarios, that we coin
{\tt F3S\_uR}, {\tt S3M\_uR} and {\tt F3V\_uR}, and that are defined in
ref.~\cite{Arina:2020udz}. Their common features are the following. First, the
dark matter candidate is a real particle, singlet under the SM gauge group, so
that its stability can be ensured through a $\mathbb{Z}_2$ symmetry. This
contrasts with
other $t$-channel models including a complex dark matter field and thus
exhibiting instead a continuous unbroken global $U(1)$ symmetry. Second, the
mediator
couples the dark matter candidate to the right-handed up-quark field, so that
the mediator is itself an $SU(2)_L$ weak singlet.
Other choices are however possible, as dark matter could interact with different quark flavours and chiralities. The $u_R$ choice is only one of the numerous possibilities, justified by its simplicity (it only involves weak singlets) and by the enhancement of the relevant collider and direct detection processes due to valence quarks. We will nevertheless highlight, in the following, wherever other mediator choices could make a difference. The defining features of the
three scenarios then consist in the spins of the dark matter and of the
mediator, which affect the kinematics of any signal and therefore current bounds
and projections for future searches. We comprehensively derive updated
constraints on the three model parameter spaces, considering both cosmological
and collider observations. Moreover, we additionally provide projections for the
future high-luminosity phase of the LHC (HL-LHC).

The rest of the paper is organised as follows. In the next section we briefly
define the \dmsimpt\ general framework for $t$-channel dark matter models, while
in~\cref{sec:lhc} we describe our analysis of the collider constraints and
provide results with current exclusion bounds. In~\cref{sec:cosmo} we study the astrophysical and cosmological constraints on these simplified models under the assumption of thermal relic dark matter.
In~\cref{sec:combined} we combine these results and include future experiment
expectations, illustrating the impact of the collider/cosmology combination on
representative projections of the model parameter space. We summarise our main findings and discuss future developments in~\cref{sec:conclusions}.

\section{The $t$-channel simplified models}\label{sec:model}

The three simplified models under study are defined with\-in the \dmsimpt\
framework~\cite{Arina:2020udz}, which provides a generic $t$-channel dark matter
simplified model. In the latter, the SM is extended by six real or complex dark matter fields, collectively denoted by $X$ and all singlets under the SM gauge group $SU(3)_c\times SU(2)_L\times U(1)_Y$, plus the corresponding mediator particles, collectively denoted by $Y$, all lying in the fundamental representation of $SU(3)_c$ and coupling the $X$ particles to the SM quarks.

The scenarios considered in the present analysis are restrictions of the general
\dmsimpt\ framework to setups in which the dark matter particle $X$ is real and
solely couples to the right-handed up-quark. There is hence a unique mediator
particle $Y$, singlet under $SU(2)_L$. The corresponding interaction
Lagrangians for the {\tt F3S\_uR} (real scalar dark matter $\tilde S$ with a
fermionic mediator $\psi$), {\tt S3M\_uR} (Majorana dark matter $\tilde\chi$
with a scalar mediator $\varphi$) and {\tt F3V\_uR} (real vector dark matter
$\tilde V_\mu$ with a fermionic mediator $\psi$) models respectively read
\be
\bsp
 \lag_{\tt F3S\_uR} = &\ \Big[
     {\bf \hat\lambda_{\sss \psi}} \bar\psi \uR \tilde S
   + {\rm h.c.} \Big] \ , \\
 \lag_{\tt S3M\_uR} = &\
   \Big[\lambda_{\sss\varphi} {\tilde\chi}\uR\varphi^\dag
  + {\rm h.c.} \Big] \ , \\
  \lag_{\tt F3V\_uR} = &\ \Big[
     {\bf \hat\lambda_{\sss \psi}} \bar\psi\slashed{\tilde V}\uR
   + {\rm h.c.} \Big] \ .
   \esp
\ee
In those expressions, $\hat\lambda_{\sss \psi}, \lambda_{\sss\varphi}$ and $\hat
\lambda_{\sss \psi}$ stand for real coupling strengths, that together with the
dark matter ($M_S, M_\chi$ and $M_V$) and mediator ($M_\psi,M_\varphi$ and
$M_\psi$) masses lead to three free parameters for each of the considered
models. We collectively denote this set of free parameters by $\{m_X, m_Y,
\lambda \}$.

In this work, we allow the two masses $m_X$ and $m_Y$
to vary in the $[1, 10^4]$~GeV range and consider $\lambda$ coupling values in
the $[10^{-4}, 4\pi]$ range (couplings larger than $4\pi$ are shown
in our results, but the $4\pi$ contour is always highlighted when relevant).
We use the corresponding next-to-leading-order (NLO)
UFO~\cite{Degrande:2011ua} model files with five massless quarks for collider
studies with \mg~\cite{Alwall:2014hca}, and both the lea\-ding-order (LO) UFO
and {\sc CalcHep}~\cite{Belyaev:2012qa} model files with six massive quarks for simulations with
\maddm~\cite{Ambrogi:2018jqj} and \micromegas~\cite{Belanger:2018mqt}
respectively. All those model files have been obtained with
\fr~\cite{Alloul:2013bka} and are available from
\url{https://feynrules.irmp.ucl.ac.be/wiki/DMsimpt}.

\section{Collider bounds}\label{sec:lhc}
\begin{figure*}
\centering
\includegraphics[clip,width=.3\textwidth]{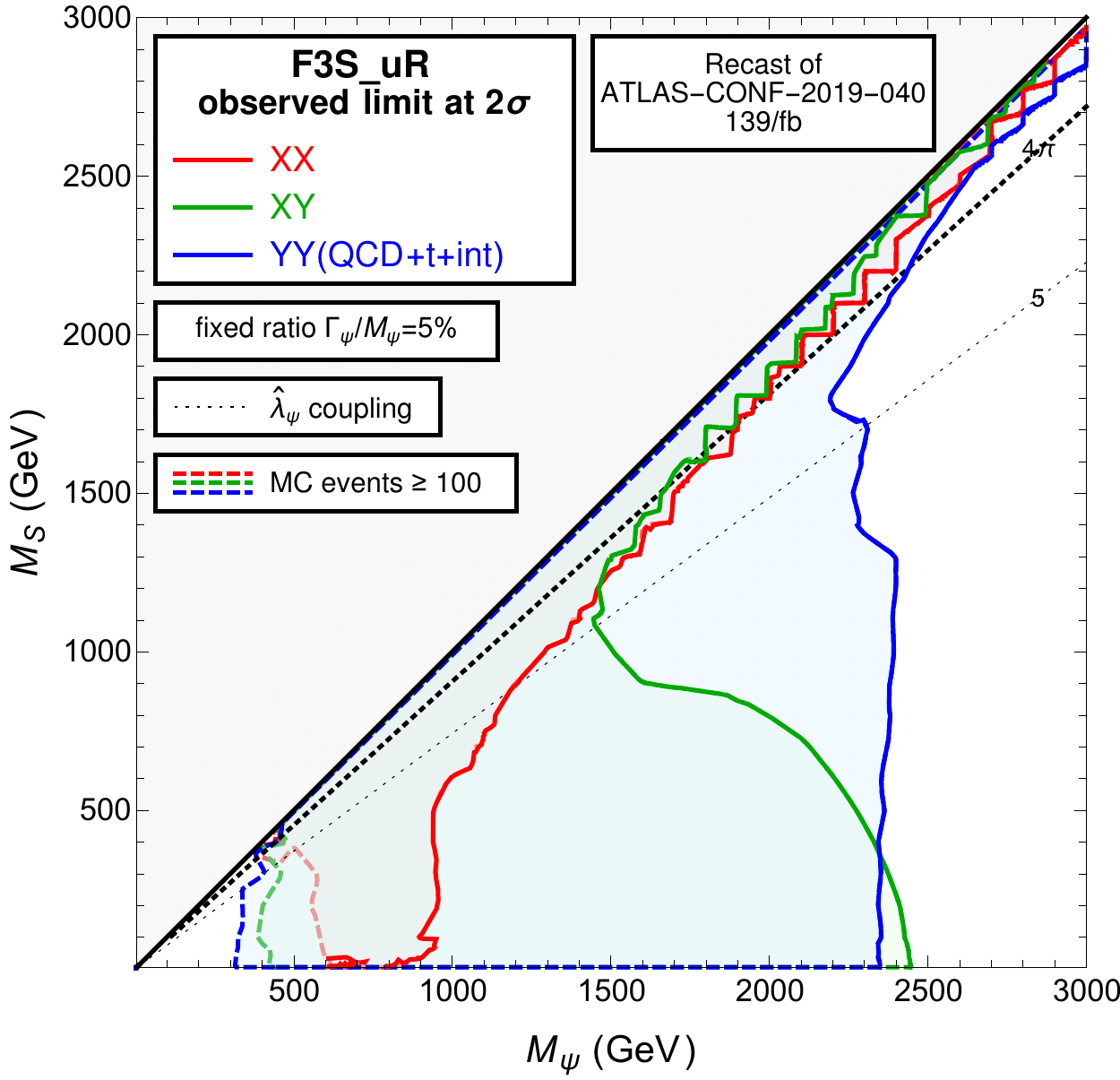}
\includegraphics[clip,width=.3\textwidth]{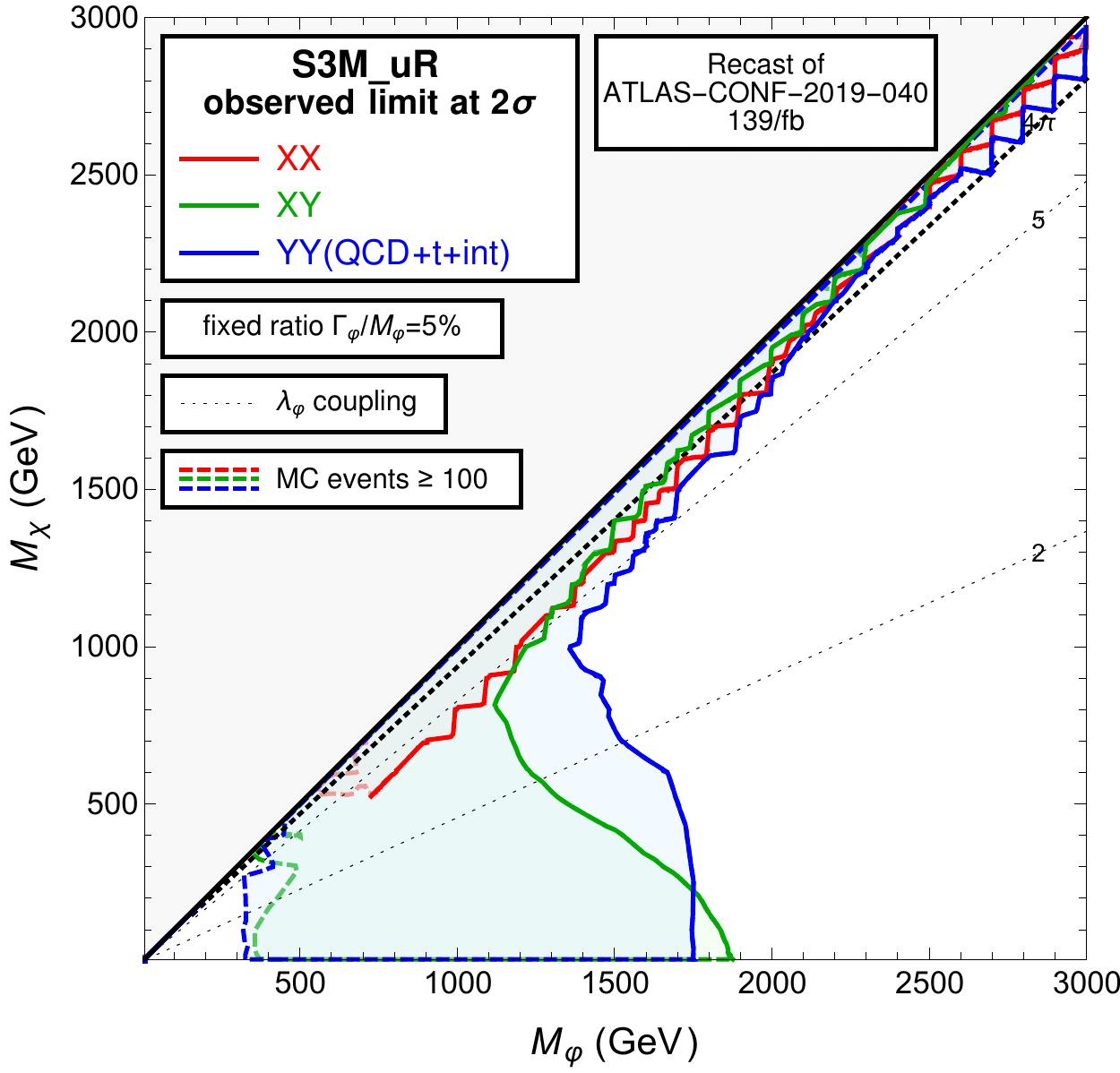}
\includegraphics[clip,width=.3\textwidth]{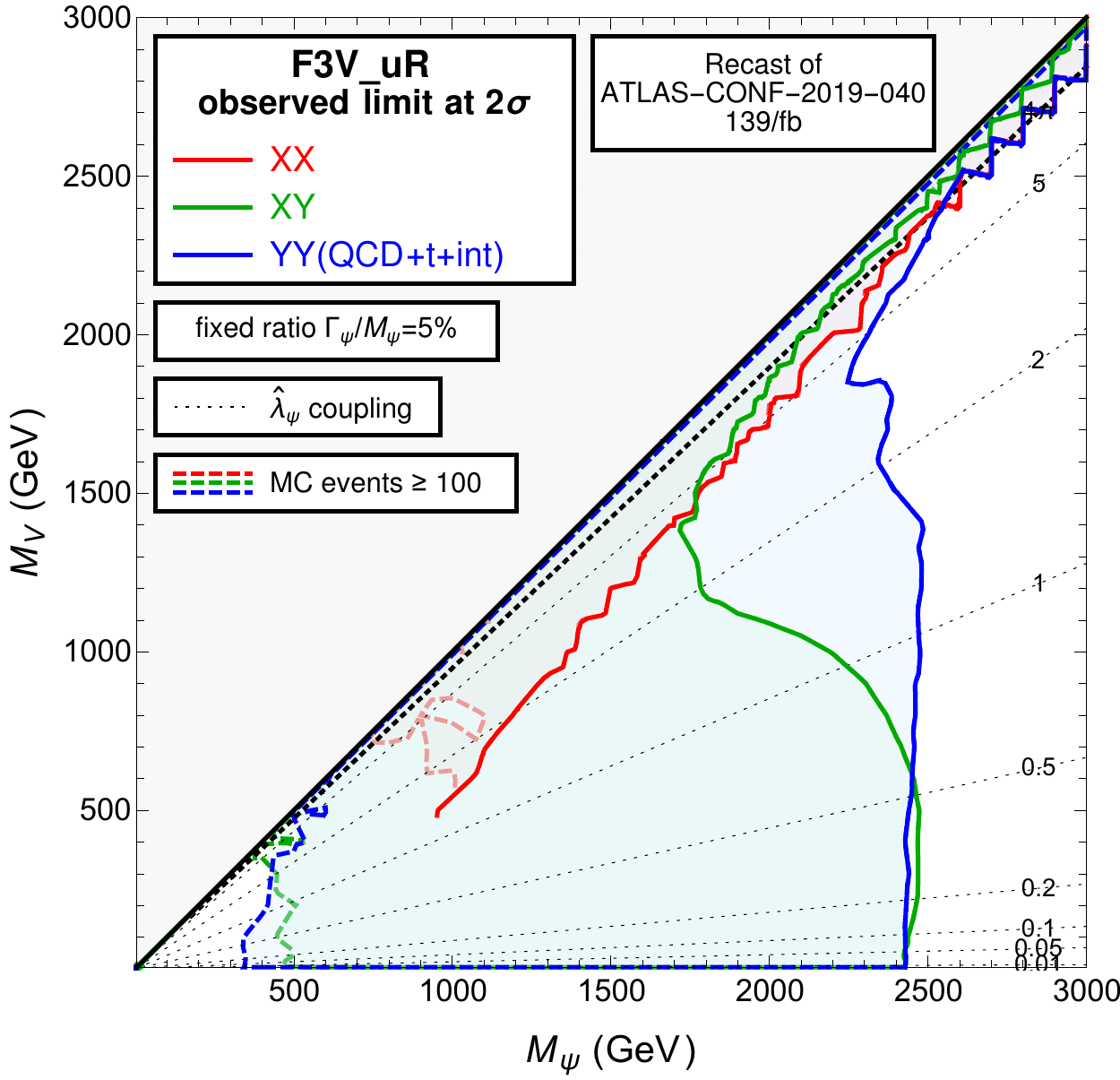}
\begin{minipage}{.08\textwidth}\quad\end{minipage}\\
\includegraphics[clip,width=.3\textwidth]{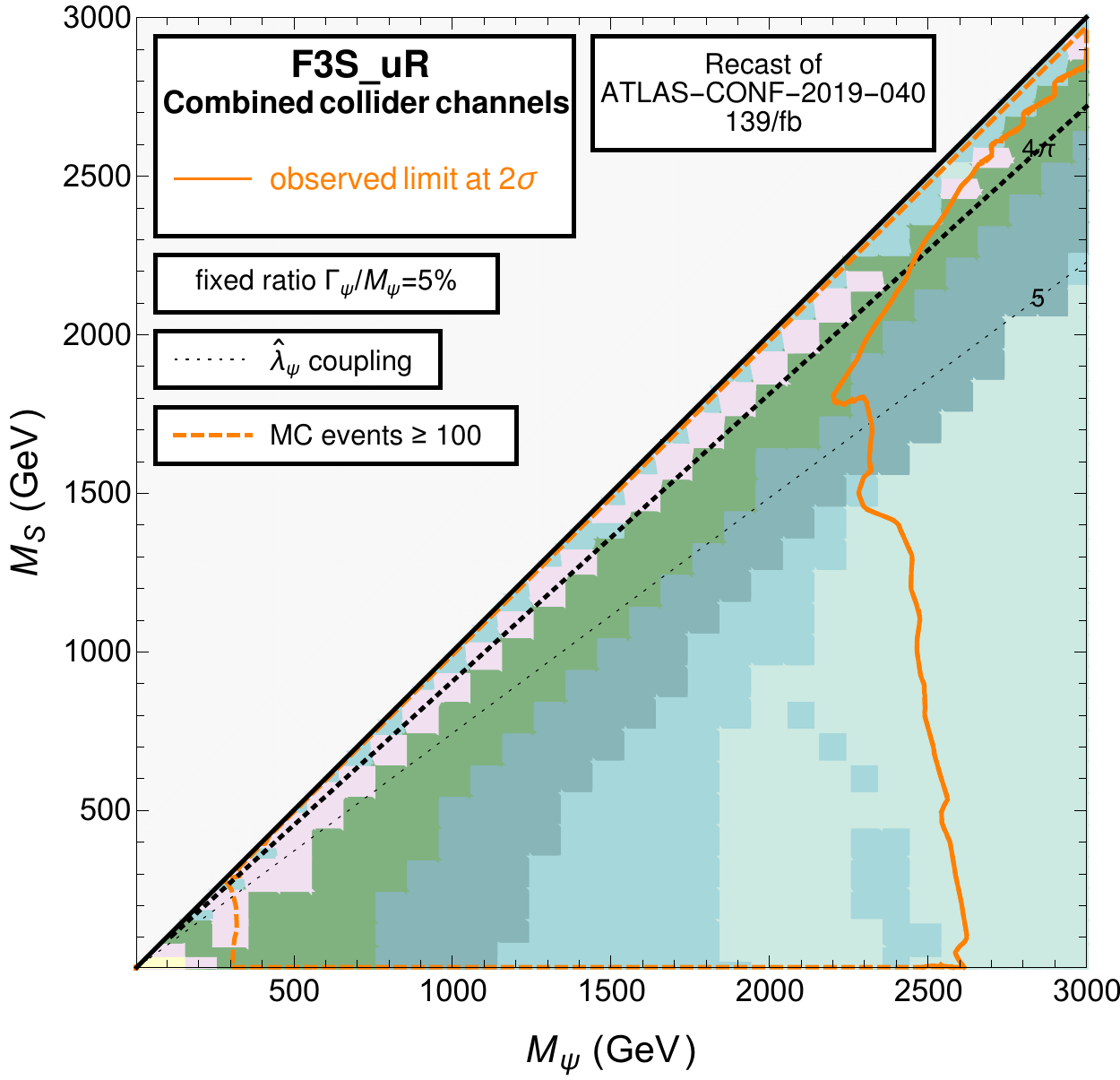}
\includegraphics[clip,width=.3\textwidth]{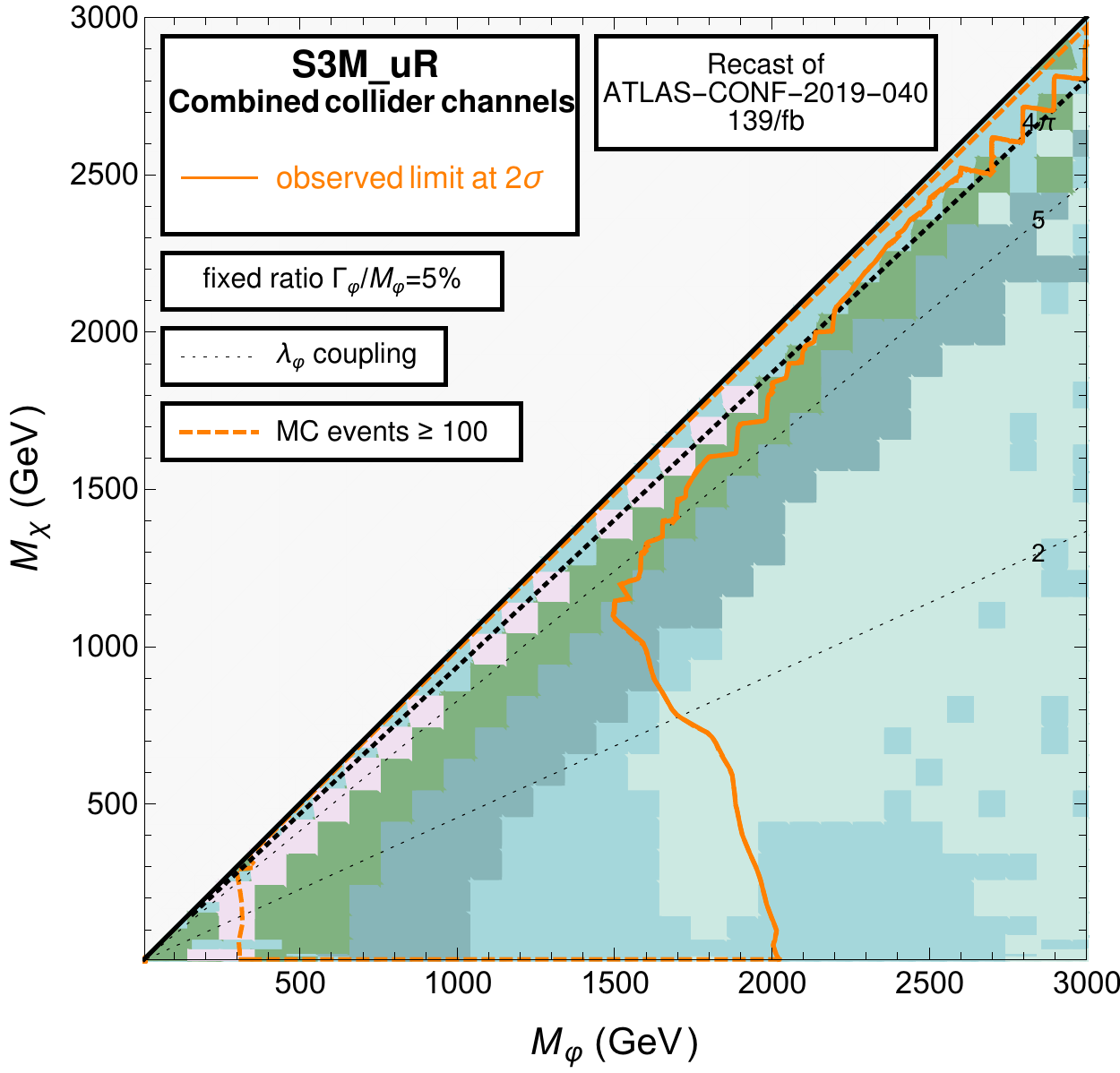}
\includegraphics[clip,width=.3\textwidth]{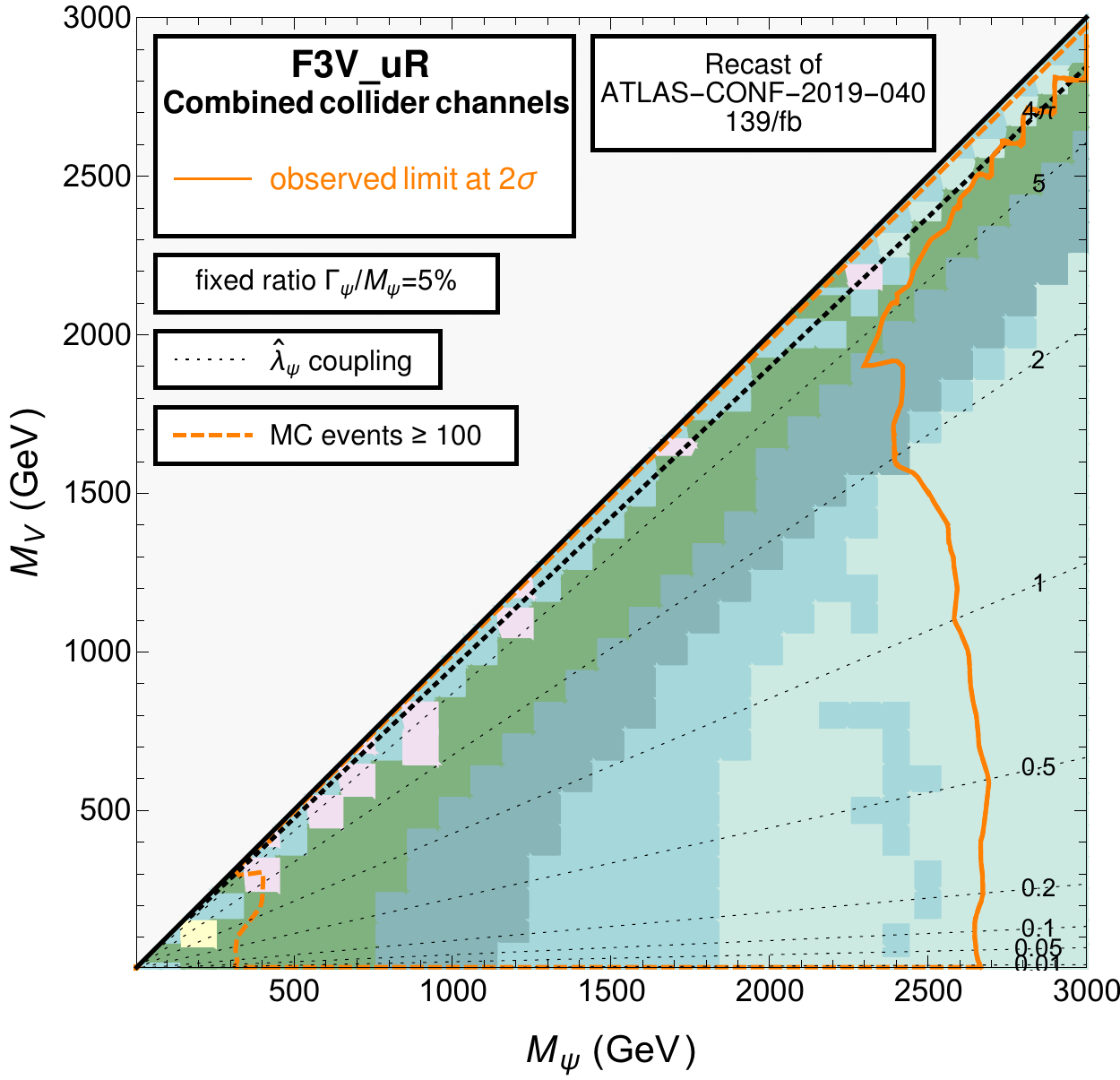}
\includegraphics[trim=0 -44pt 0 0,clip,width=.08\textwidth]{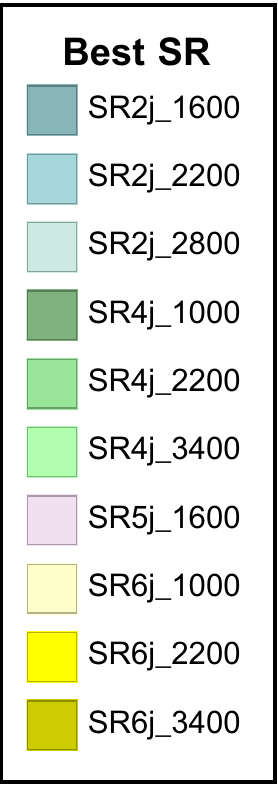}
\caption{
\textbf{Top row:} Individual 95\% CL bounds arising from the three different
channels $XX$ (red), $XY$ (green) and $YY$ (blue) for the {\tt F3S\_uR} (left),
{\tt S3M\_uR} (centre) and {\tt F3V\_uR} (right)
scenarios, presented in the $(m_{Y}, m_{X})$ plane for a fixed mediator
width-to-mass ratio. \textbf{Bottom row:} Combined 95\% CL bounds, with the
signal region exhibiting the best sensitivity depicted by the background colour.
In all panels, the black dashed lines correspond to the value of the couplings
which is required to obtain a width over mass ratio of 5\%, the 4$\pi$ value
being highlighted to roughly identify the perturbative regime. The coloured
dashed lines identify the area for which the number of simulated Monte Carlo
events populating the best region is larger than 100 (allowing for a Poisson
uncertainty smaller than 10\%).}
\label{fig:colliders}
\end{figure*}

Three types of processes are considered for the determination of the collider
constraints on the models. They consist in the production of a pair of dark
matter particles ($p p \to X X$), of a pair of mediators ($p p \to Y Y$) and the
associate production of a dark matter and a mediator ($p p \to X Y$).
Mediator pair-production is itself composed of three components, namely a QCD
contribution, a dark-matter-induced contribution (with the propagation of the
dark matter particle in the $t$-channel) and their interference. When the
mediator is produced, it subsequently decays into a dark matter candidate and
a right-handed up-quark ($Y \to X u_R$), the decay process being always
factorised from the production one. This however assumes that the decay width
of the mediator $\Gamma_Y$ is small relatively to its mass, such that the
narrow-width
approximation (NWA) holds\footnote{In principle, the NWA approximation is valid
only if the mass difference between the decaying particle and the decay products is large~\cite{Berdine:2007uv}. We however assume that corrections arising from
small mass splittings are not significant in the corresponding regions of the parameter space.}.
The relative contributions of the different channels depend on the
exact details of the model, and in particular on the $\lambda$ coupling value.
In particular, the relevance of the $XX$ channel originates from the emission of jets by the initial state and the internal mediator, that are both considered at the matrix-element and parton-shower level in our NLO simulations matched with parton showers.
In terms of the kinematics, the channels with the largest cross section are,
however, not necessarily the relevant
ones in terms of probing the model parameter space and setting limits, as
already illustrated in ref.~\cite{Arina:2020udz}.

All simulations are performed with \mg\ and follow the procedure described in
ref.~\cite{Arina:2020udz}, the NLO matrix elements being convoluted with the
NNPDF~3.0 set of parton densities~\cite{Ball:2014uwa} through the {\sc LHAPDF 6}
library~\cite{Buckley:2014ana}. Moreover, to ensure the validity of the NWA and
the factorisation of the production and decay processes, all simulations have
been performed at a fixed $\Gamma_Y/m_Y$ ratio of 1\%, assuming that the
final-state kinematics is not impacted by slightly larger values of this ratio.
In the following, we reweigh those generated events so that the cross section
evaluation makes use of a $\lambda$ value yielding $\Gamma_Y/m_Y = 5\%$. This
choice requires a more important coupling and leads to weaker cosmological
constraints, which thus allows for a larger cosmologically-viable region of the
parameter space to be probed by LHC searches (see \cref{sec:cosmo} and
\cref{sec:combined}).
Different choices of the $\Gamma_Y/m_Y$ ratio can impact the
results, as lower $\Gamma_Y/m_Y$ values imply lower couplings. Besides an
expected strengthening of the relic density constraints yielded by a smaller
annihilation cross section, it would reduce the relative
weights of the $XX$, $XY$ and non-QCD $YY$ collider channels with respect to QCD
$YY$ production whose cross section is independent of $\lambda$. On the other
hand, larger width-over-mass ratios would make the collider analysis less
accurate, as for large widths the NWA-motivated factorisation of the production
and decay processes would not accurately describe the kinematics of the final state.

We obtain constraints on the models through the recast of an ATLAS search
targeting final states with multiple jets and missing transverse
energy~\cite{ATLAS:2019vcq} by means of the \ma\ framework~\cite{Conte:2018vmg,ATLASCONF2019040recast}.
This search is well suited to probe scenarios where dark matter
interacts with light quark flavours, as considered in this work. While monojet
searches could be relevant too, they consist strictly speaking in multijet plus
missing energy searches, as a subleading jet activity is allowed. They are thus
only different from the considered search by the details of the requirements on
the event hadronic activity. As no ATLAS and CMS monojet search has
been updated as a full run~2 analysis yet, monojet probes will be ignored.
Other searches
could nevertheless be better in other contexts. For instance, for setups
involving interactions with top quarks, searches involving final-state top quarks and missing transverse energy could probably give a slightly better reach, as already found out for instance for scalar~\cite{Colucci:2018vxz} or Majorana~\cite{Garny:2018icg} dark matter.
The significance of the signal is derived for each of the ten signal regions
(SRs) of the search through the CLs method~\cite{Read:2002hq}, and we include
in our predictions signal systematics stemming from scale variations and the
parton density fits~\cite{Araz:2019otb}.
The yields of the backgrounds for each SR, with their uncertainties, and the number of observed events, are provided by the ATLAS search.
As observations are compatible with the background within 1$\sigma$
for all signal regions of relevance, observed and expected
bounds do not significantly differ. We show the former in our results.
Obviously, for high-luminosity projections (see section~\ref{sec:combined}),
expected limits are used for the extrapolations.

Due to the different dependence of the cross sections on the $\lambda$ coupling
and on the masses of the new particles, the relative weights of the $XX$, $XY$
and $YY$ contributions in the determination of the constraints change along the
parameter space, as shown in the top row of~\cref{fig:colliders}.
The combination of the various contributions to the $YY$ process constrains the majority of the parameter space for all the considered scenarios.
In contrast, the $XX$ process only becomes competitive in the compressed region and for large mediator masses, while the $XY$ one provides instead stronger constraints for scenarios featuring a large mass gap and a large mediator mass.
The region with small dark matter and mediator masses is likely to be excluded too, but the number of initial MC events required to test the region with enough statistics is too demanding in terms of computing resources.

The combination of the bounds for any given scenario is obtained in two steps.
We first sum the number of events populating each signal region as obtained from
the individual $XX$, $XY$ and $YY$ contributions, and then compute the
corresponding significance. We display the results in the bottom row
of~\cref{fig:colliders},
in which we additionally highlight the best signal region driving the bound.
The dominance of the $YY$ component in the determination of the bounds is reflected in the similarities of the results for the {\tt F3S\_uR} and {\tt F3V\_uR}
models that share the same mediator particle. For the {\tt S3M\_uR} class of
scenarios, the bounds are sizeably weaker, given the smaller cross section for
the pair production of a scalar mediator that features a smaller number of
degrees of freedom than a fermion.

The combined signal kinematics for any given scenario depends on the subprocess
that dominates, which is reflected in the variations in the best SRs driving the
bounds along the parameter space. Regions requiring two very hard jets are more
suitable when the $YY$ channel dominates and each mediator decay leads to a
significantly hard jet. In contrast, SRs dedicated to final states featuring
four jets give a better outcome in the compressed regime. While these regions
select events exhibiting a larger number of jets, the associated transverse
momentum requirements are milder than in the two-jet case, and thus more
efficient in more compressed setups in which decay and radiation jets are
softer.

\section{Cosmological bounds}\label{sec:cosmo}
\begin{figure*}
\centering
\includegraphics[trim=25. 20. 55. 53.,clip,width=.32\textwidth]{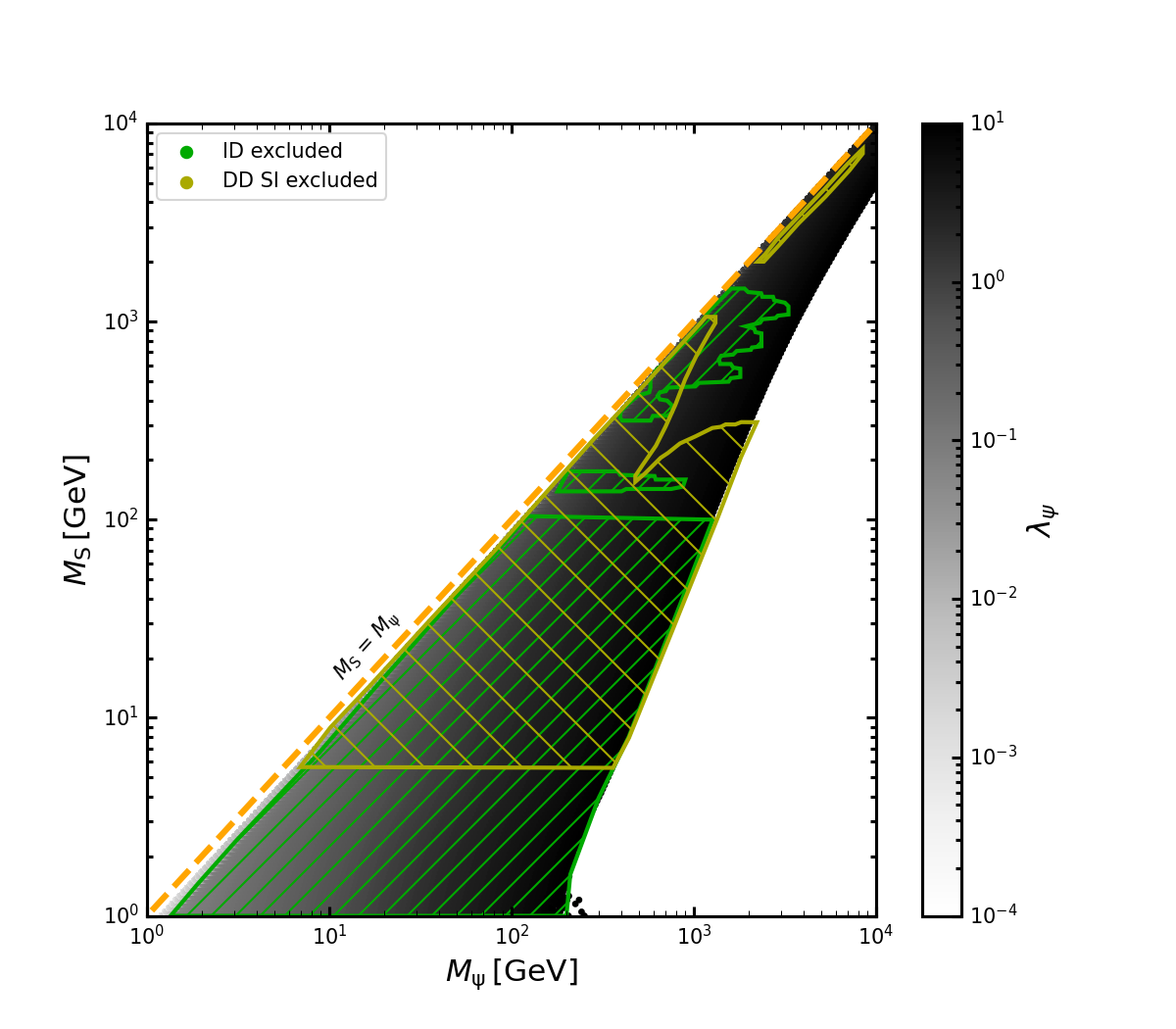}
\includegraphics[trim=25. 20. 55. 53.,clip,width=.32\textwidth]{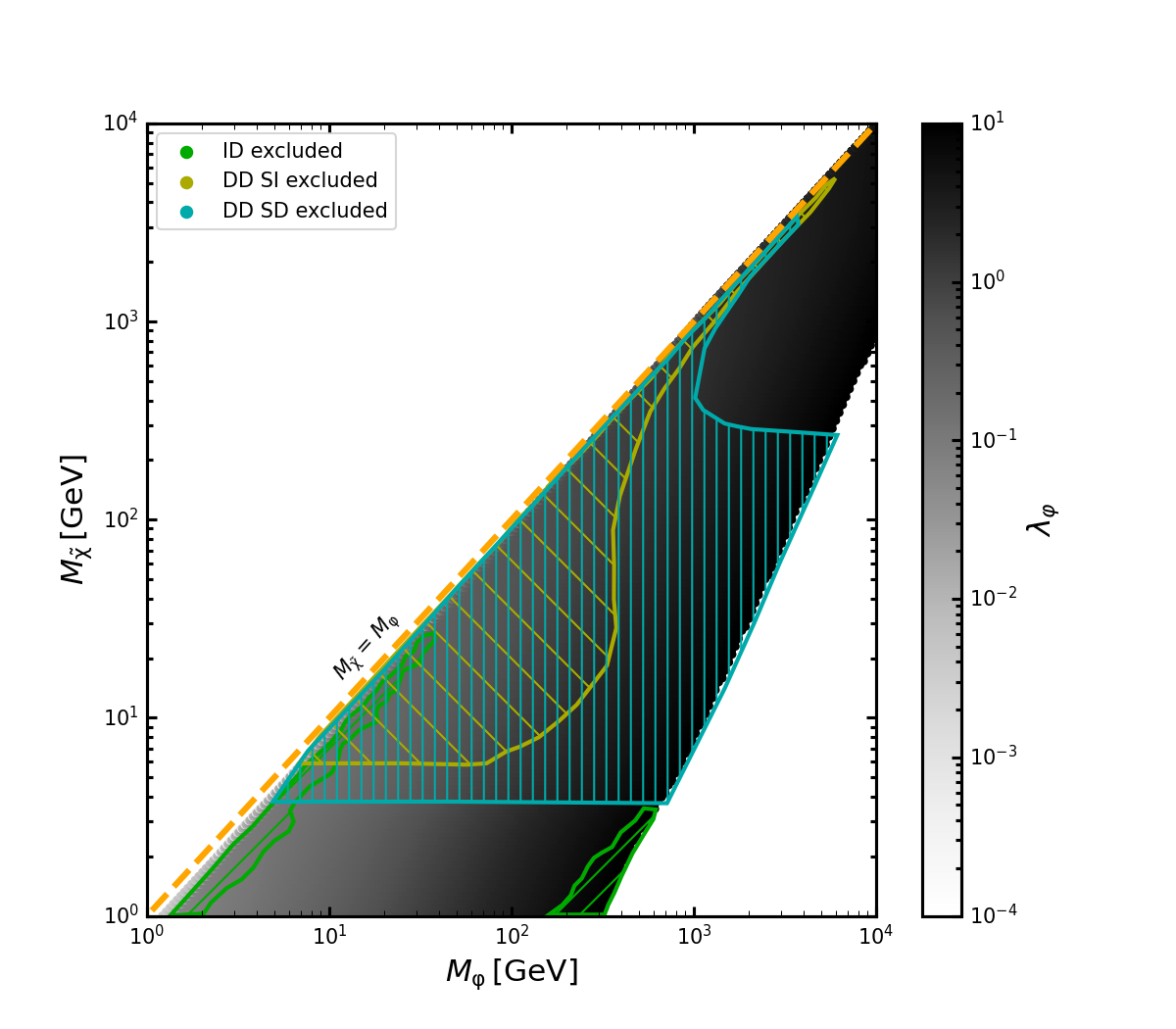}
\includegraphics[trim=25. 20. 55. 53.,clip,width=.32\textwidth]{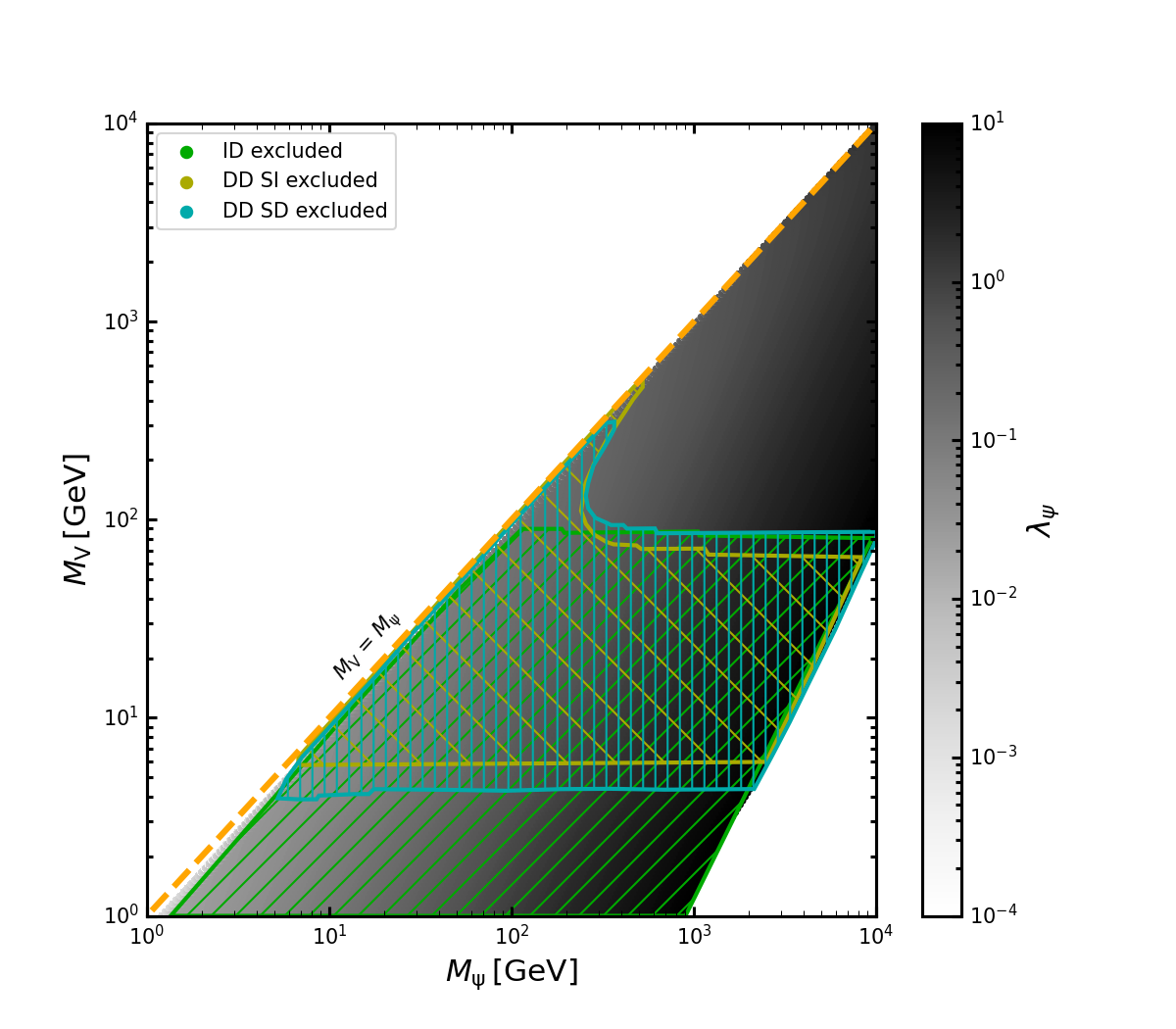}\\
\includegraphics[trim=25. 20. 55. 53.,clip,width=.32\textwidth]{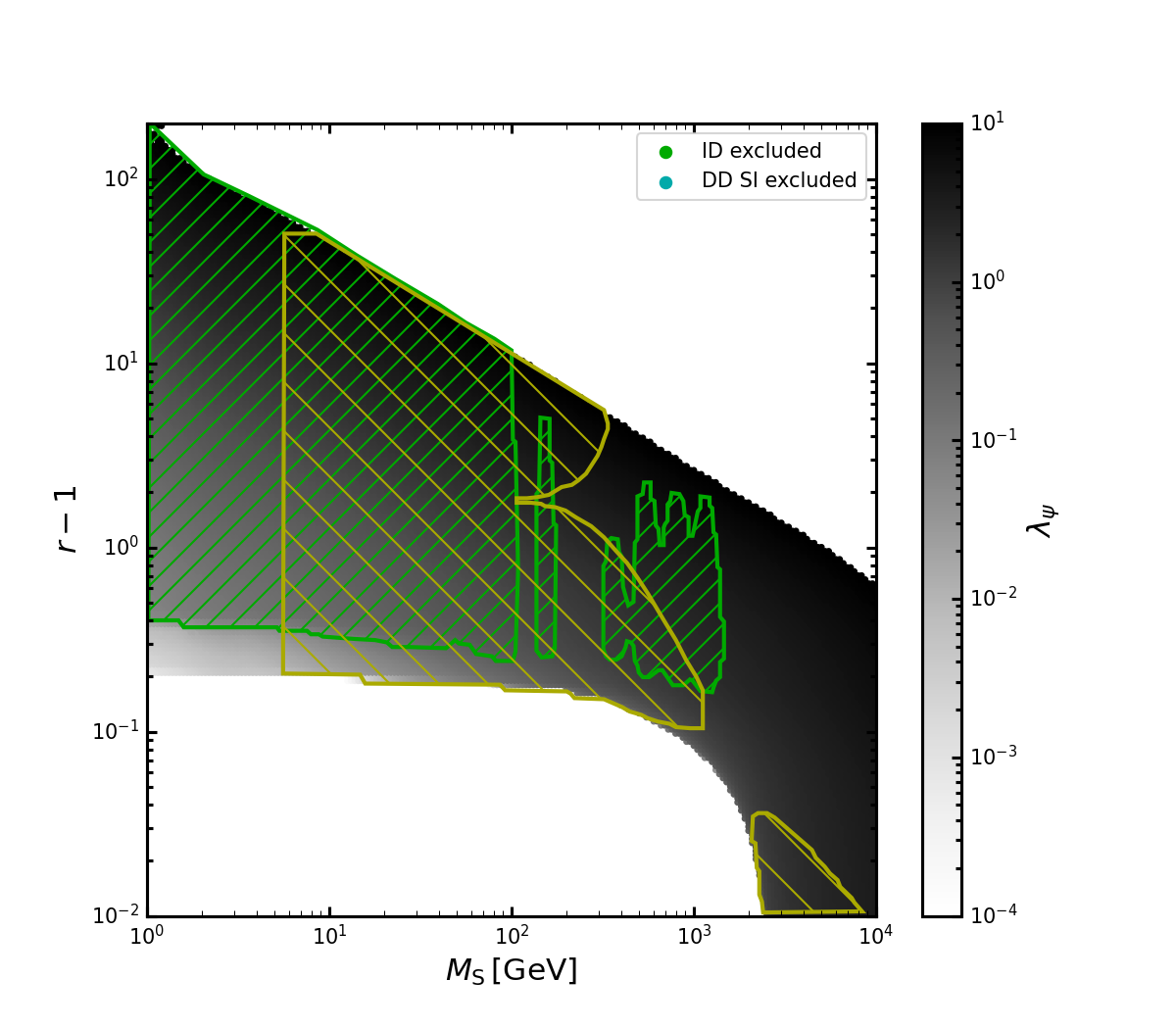}
\includegraphics[trim=25. 20. 55. 53.,clip,width=.32\textwidth]{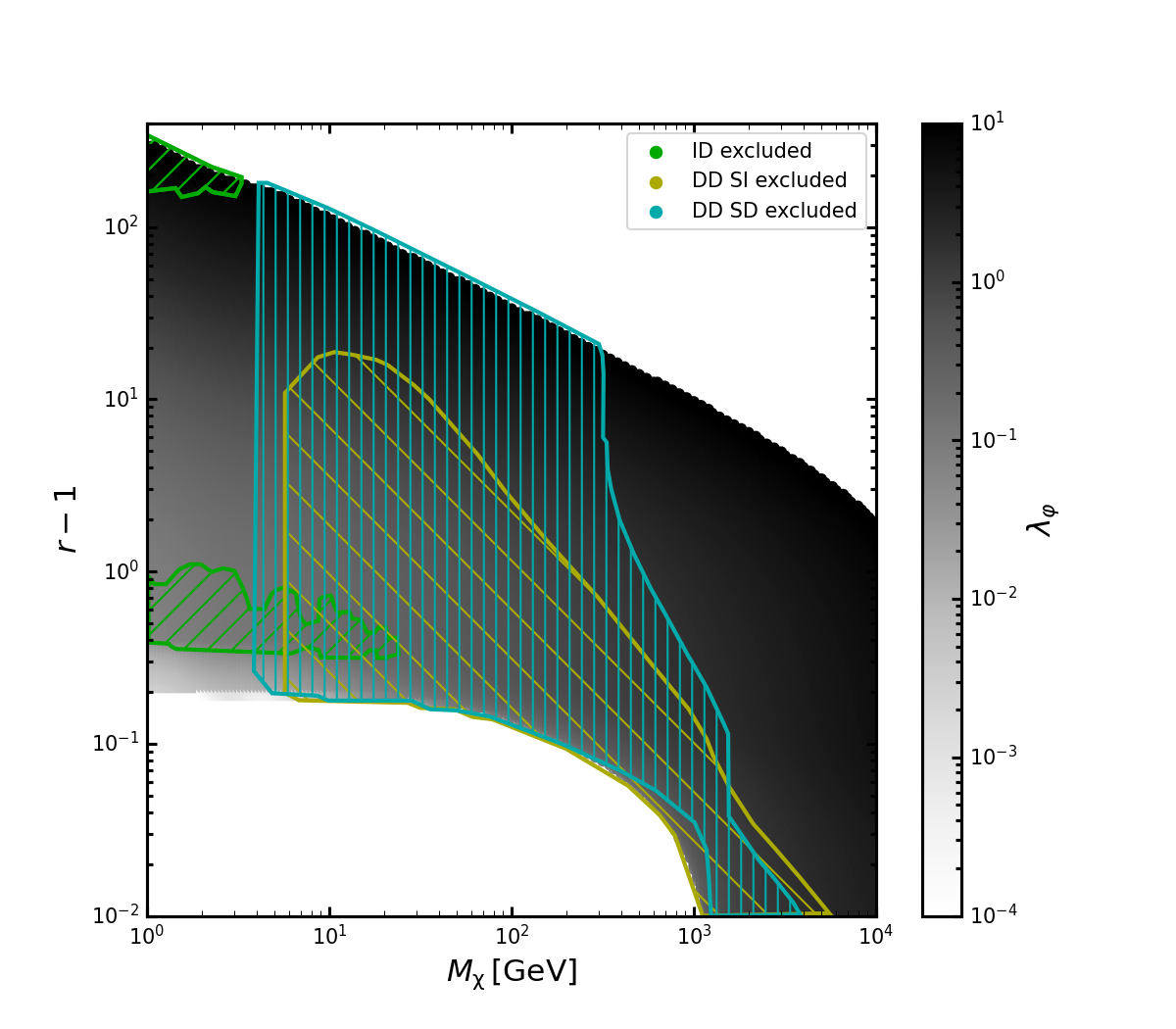}
\includegraphics[trim=25. 20. 55. 53.,clip,width=.32\textwidth]{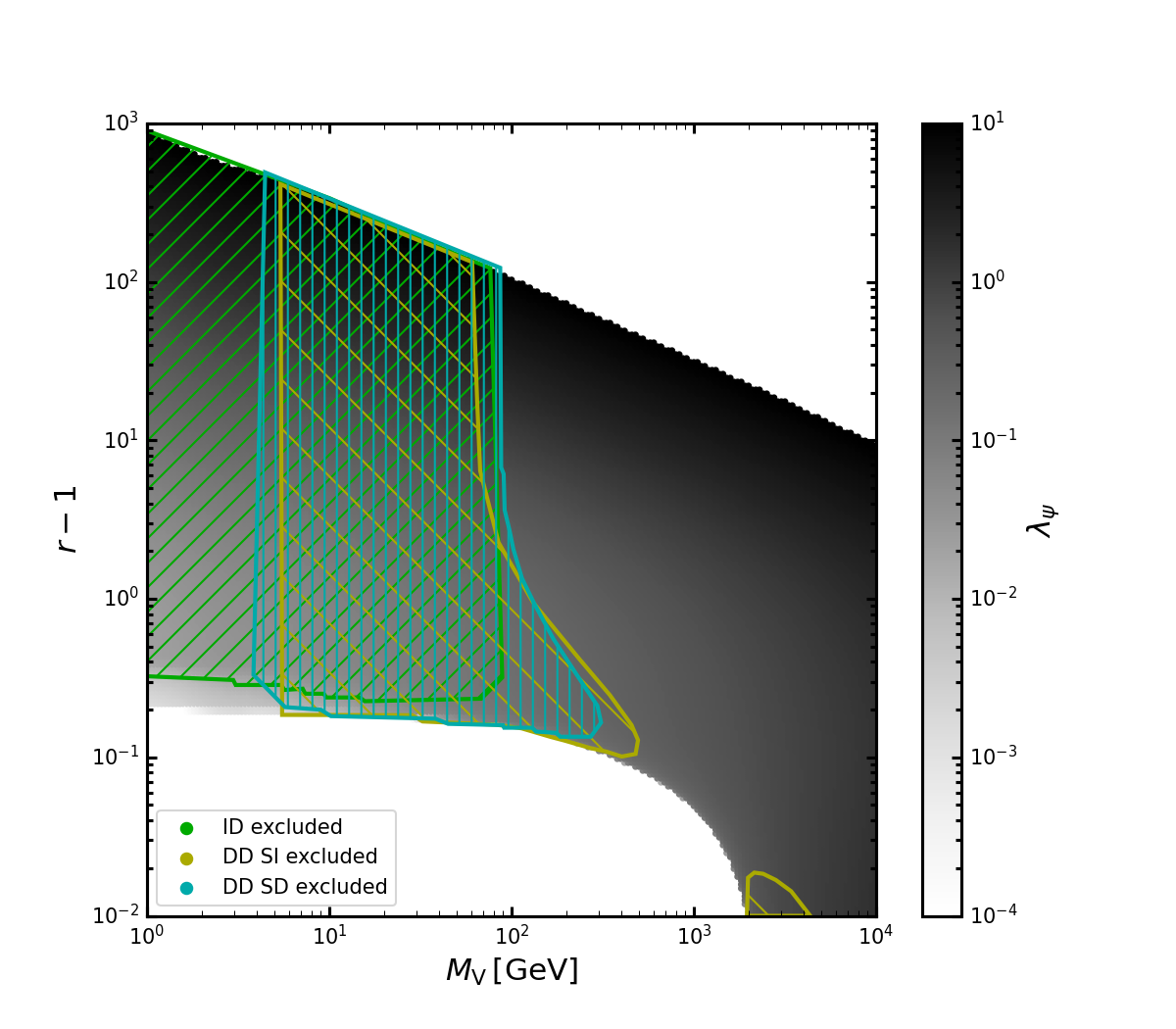}
\caption{ {\bfseries Top row}: Parameter space regions compatible with the
standard freeze-out mechanism and the observed relic
density~\cite{Aghanim:2018eyx}, shown in the $(m_{Y}, m_{X})$ plane for the
{\tt F3S\_uR} (left), {\tt S3M\_uR} (centre) and {\tt F3V\_uR} (right) models.
The gray shading indicates the $\lambda$ value needed to satisfy the relic
density constraint. The yellow hatched region is excluded by the SI XENON1T
bounds~\cite{Aprile:2018dbl} (DD SI), the green hatched one by gamma-ray line searches
from Fermi-LAT~\cite{Ackermann:2015lka} and HESS~\cite{Abdallah:2018qtu} (ID), and
the cyan hatched one by the SD PICO bounds~\cite{Amole:2017dex} (DD SD). Moreover, for
the {\tt F3V\_uR} model, indirect detection bounds are extracted from
Fermi-LAT in dSphs searches~\cite{Fermi-LAT:2016uux} (ID), when one relies on
dark matter annihilations in the $u\bar u$ final state~\cite{Ambrogi:2018jqj}.
\textbf{Bottom row}: Same as for the top row but in the $(r-1, m_{X})$ plane
where $r \equiv m_{Y} / m_{X}$. This allows us to highlight better the
co-annihilation regime.}
\label{fig:cosmo_combined}
\end{figure*}

For all three models, we sample the three-dimensional parameter space with
\micromegas\ and require that the dark matter candidate makes up $100\%$ of the
measured dark matter abundance, $\Omega h^2_{\rm Planck}=0.12$~\cite{Aghanim:2018eyx}.
The thermally averaged dark matter annihilation cross section $\langle\sigma v
\rangle$ ($v$ being the relative velocity between two dark matter particles) is
$d$-wave-suppressed for the real scalar case~\cite{Toma:2013bka,Giacchino:2013bta,
Giacchino:2014moa,Giacchino:2015hvk,Biondini:2019int} and $p$-wave-suppressed for Majorana dark
matter~\cite{Giacchino:2014moa,Biondini:2018ovz}. NLO corrections in the relic density
computation might therefore be relevant~\cite{Giacchino:2015hvk,
Colucci:2018vxz}. To account for
these corrections, we include the loop-induced $X X \to g g$ and $X X\to\gamma
\gamma$ processes\footnote{$X X \to \gamma Z$ annihilations should be included
as well, as the associated matrix element is of the same perturbative order as
the $XX\to\gamma\gamma$ one. However, we have found out that the di-photon
contribution to $\langle\sigma v\rangle$ is subdominant to the $XX\to
gg$ one in the entire parameter space. We have therefore not accounted
for annihilations into  a $\gamma Z$ system, that is itself subleading
with respect to $XX\to\gamma \gamma$.},
and the three-body $X X \to \uR \uRb g$ and $X X \to \uR \uRb \gamma$
annihilations that could be potentially enhanced by virtual
internal bremsstrahlung (VIB). For our predictions, we use the analytic
expressions provided in refs.~\cite{Giacchino:2013bta,Giacchino:2014moa,
Ibarra:2014qma} that we have validated with \maddm.
While different choices of dark matter interactions (in terms
of the flavour and chirality of the involved SM quarks) would lead to a different
interplay between the subprocesses contributing to the relic density, it will always be
possible to find viable solutions for the $\lambda$ parameter.

Through our scans of the model parameter spaces, we single out regions
where the elastic dark matter scattering cross section off protons is compatible
with both the spin-independent (SI) and spin-dependent (SD) exclusion limits at
90\% confidence level (CL) from the XENON1T~\cite{Aprile:2018dbl} and
PICO~\cite{Amole:2017dex} experiments, our predictions relying on NLO cross
sections~\cite{Hisano:2015bma} to properly model the impact of QCD radiation.
In principle, running coupling effect should also be
included~\cite{Mohan:2019zrk}. The latter would lead to tighter exclusion
limits, slightly augmenting their sensitivity for large dark matter masses. We
have however omitted them from our computations, although we have verified that
they do not impact our conclusions. We do not expect the obtained
direct detection bounds to sensibly change for different choices of dark matter
interactions with right-handed valence quarks. For other scenarios involving sea
quarks, we however expect those bounds to be weakened.
Finally, we impose in our scanning procedure that predicted indirect
detection signals are compatible with the current (model-dependent) exclusion
limits at 95\% CL. This time, different choices of dark matter
interactions would result in a different weighting of the subprocesses
contributing to the gamma-ray signals, and therefore of different results.

In the case of the {\tt F3S\_uR} and {\tt S3M\_uR} models, spectral features in
the gamma ray spectrum bring one of the strongest bounds as tree-level $X X \to
\uR \uRb$ annihilations are velocity suppressed. We therefore derive constraints
by considering a combination of direct annihilations into photons and into a
$\uR\uRb\gamma$ system, the latter being potentially enhanced by VIB
contributions. The total annihilation cross section $\langle\sigma v\rangle_{\rm
tot} = \langle\sigma v\rangle_{\uR\uRb\gamma}+2\langle\sigma v\rangle_{\gamma
\gamma}$ is then confronted with the most recent
Fermi-LAT~\cite{Ackermann:2015lka} and HESS~\cite{Abdallah:2018qtu} data from
the Galactic Centre\footnote{We derive exclusion limits by considering an
Einasto dark matter density profile~\cite{Einasto:2009zd}.}. We assume that the gamma-ray spectrum related to
the $\uR\uRb\gamma$ contribution presents a sharp feature close to the dark
matter mass, even though the exact position of this feature depends on $r\equiv
m_{Y}/m_X$~\cite{Garny:2013ama}. The obtained constraints are in the worst case
conservative, although for most scanned over scenarios they consist in a good
approximation. The three-body signal indeed dominates over the di-photon one, at
least at small $r$ values, so that the peak is often very close to the dark
matter mass. The derivation of more precise constraints would require a recast
of the experimental results, which lies beyond the scope of this study.

Other relevant bounds can be obtained by investigating dark
matter annihilations into gluons, as this could be constrained by the Fermi-LAT
analysis of dwarf spheroidal galaxies (dSphs) data~\cite{Fermi-LAT:2016uux}.
Similarly to the gamma-ray case, we evaluate $\langle\sigma v
\rangle_{\rm tot} = \langle \sigma v\rangle_{\uR \uRb g} + \langle\sigma v
\rangle_{gg}$ and compare our predictions with Fermi-LAT dSph results for the
$gg$ annihilation channel~\cite{Ambrogi:2018jqj}. These constraints being
comparable with those arising from gamma-ray line searches, they are omitted
from the discussion. Finally for the {\tt F3V\_uR} model, $XX\to\uR\uRb$
annihilations occur in an $s$-wave configuration. The most stringent indirect
detection bounds are thus given by Fermi-LAT dSph searches, this time in the
$u\bar u$ final state.

Our results are shown in~\cref{fig:cosmo_combined}. The gray shaded region
represents scenarios that can account for the correct relic density when
assuming a standard freeze-out mechanism. For the {\tt F3S\_uR} model, NLO
corrections drastically modify the contours of the viable parameter space region
at large $r$, selecting $\lambda$ values smaller than for the LO case. This
stems from the $XX\to gg$ contributions, that are driven by the strong coupling
constant $\alpha_s$ and that enhance the annihilation cross section. On the
contrary, NLO corrections for the Majorana dark matter case do not impact the
results much. In the large $r$ regime, we obtain deviations in the $\lambda$
value of at most 15\% with respect to the LO case, whilst scenarios featuring a
small $r$ value are unaffected, the annihilation cross section being
dominated by $\alpha_s$-dependent co-annihilations. Following the same
reasoning, it turns out that the actual value of $\lambda$ is irrelevant when
co-annihilations of the mediator via QCD processes drive the relic density.

The {\tt F3V\_uR} model is the one that features the largest parameter space
for which the relic density as measured by the Planck collaboration can be
accommodated. For any given $(m_X, m_Y)$ mass configuration, the $\lambda$
value that is needed to obtain $\Omega h^2_{\rm Planck}$ is smaller than in the scalar
and Majorana dark matter cases. The annihilation strength of vector dark matter
is indeed larger, except in the co-annihilation regime where the model is
indistinguishable from the {\tt F3S\_uR} setup that also features a fermionic
mediator.

Our findings show a nice complementarity between direct and indirect dark matter
searches in the case of the {\tt F3S\_uR} and {\tt F3V\_uR} models. Gamma-ray
searches (green hatched region) are able to probe and disfavour at 95\% CL dark
matter candidates with masses ranging down to 1~GeV, except for compressed
spectra with $r-1 \lesssim 0.3$ and very small couplings below about $10^{-2}$
(bottom row of \cref{fig:cosmo_combined}). This unexplored region consists in
one of the two co-annihilation-dominated regions which are still open and might
give rise to interesting LHC signatures through long lived particles
(LLPs)~\cite{Alimena:2019zri}. For the {\tt S3M\_uR} model, indirect detection
plays a minor role, excluding a limited part of the parameter space where dark
matter is light. The two separated excluded regions correspond to Fermi-LAT
limits arising from $XX\to\gamma\gamma$ (large $r$ values) and $XX\to\uR\uRb
\gamma$ (small $r$ values) annihilations respectively. Finally, {\tt F3S\_uR}
scenarios can be proved by the HESS experiment, as depicted by the disfavoured
island in the parameter space at $m_X>300$~GeV.

Direct and indirect detection bounds both exclude the intermediate mass range,  
although direct detection bounds additionally contribute to cut down the
parameter space. This is particularly true for large dark matter
masses, close to 1~TeV or even higher,
where one finds a second viable co-annihilation regime and where
the XENON1T bounds (yellow hatched region) start to play a role. In addition,
very light dark matter scenarios are excluded due to PICO constraints (cyan hatched
region). Remarkably, these two direct
detection experiments are also able to probe the co-an\-ni\-hi\-la\-ti\-on regime. The
whole freeze out parameter space is hence disfavoured at 90\% CL for dark matter
masses between 4~GeV and 1000 (500) GeV for the {\tt F3S\_uR} ({\tt F3V\_uR})
model.  

Spin-dependent direction detection exclusion bounds are the most stringent
constraints on the
{\tt S3M\_uR} model parameter space, even though spin-independent experiments start to be sensitive to dark matter masses larger than 4 TeV. Majorana dark matter is strongly
disfavoured for masses between 8 and 300 GeV, even for the co-annihilation
regime that could give rise to LLP collider signatures. The latter regime is
even further constrained, for dark matter masses ranging up to 10~TeV, by the
XENON1T SI bounds, these constraints being due to the scalar nature of
the mediator.

\section{Combining dark matter searches}\label{sec:combined}
\begin{figure*}
\centering
\includegraphics[clip,width=.32\textwidth]{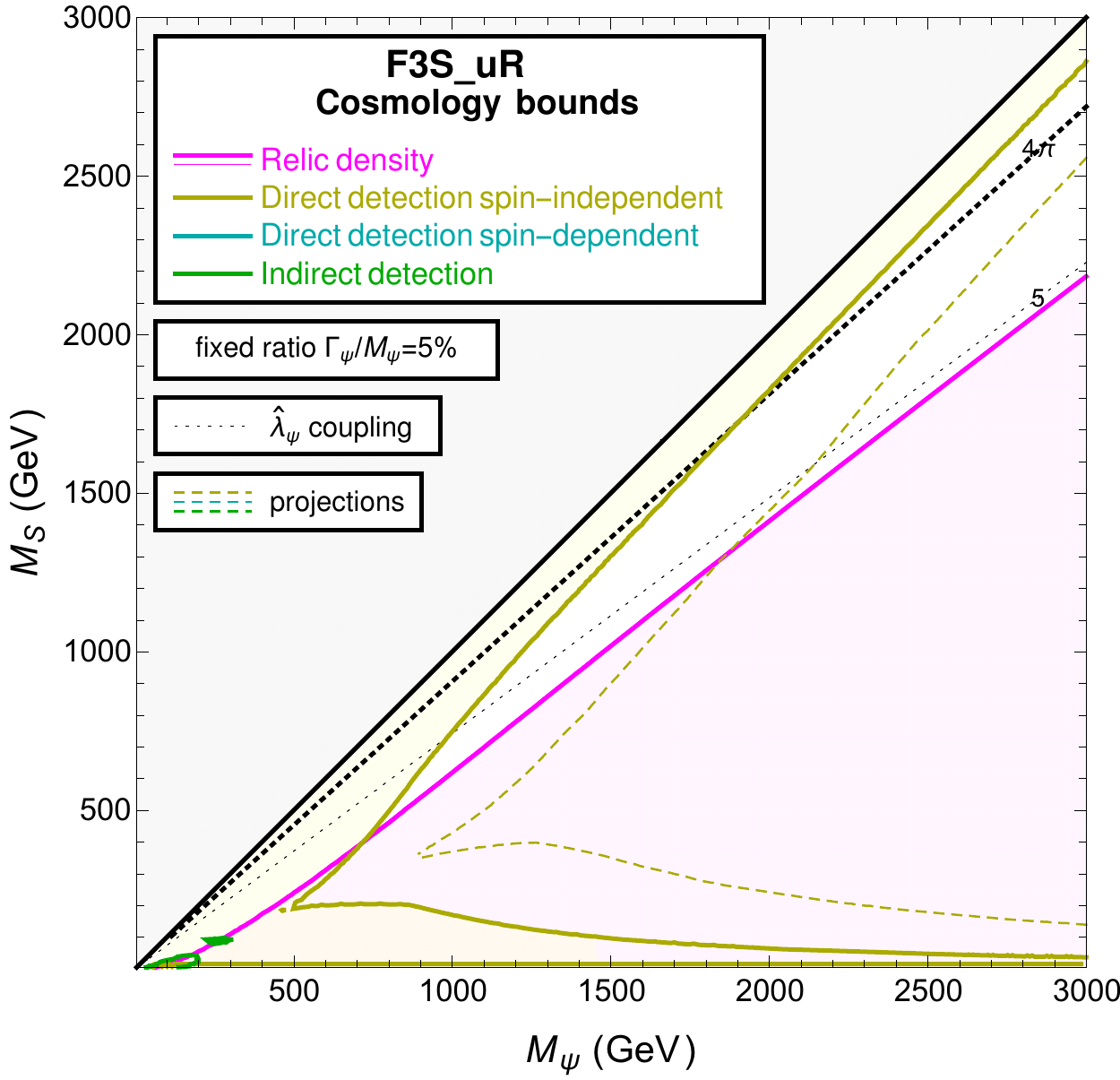}
\includegraphics[clip,width=.32\textwidth]{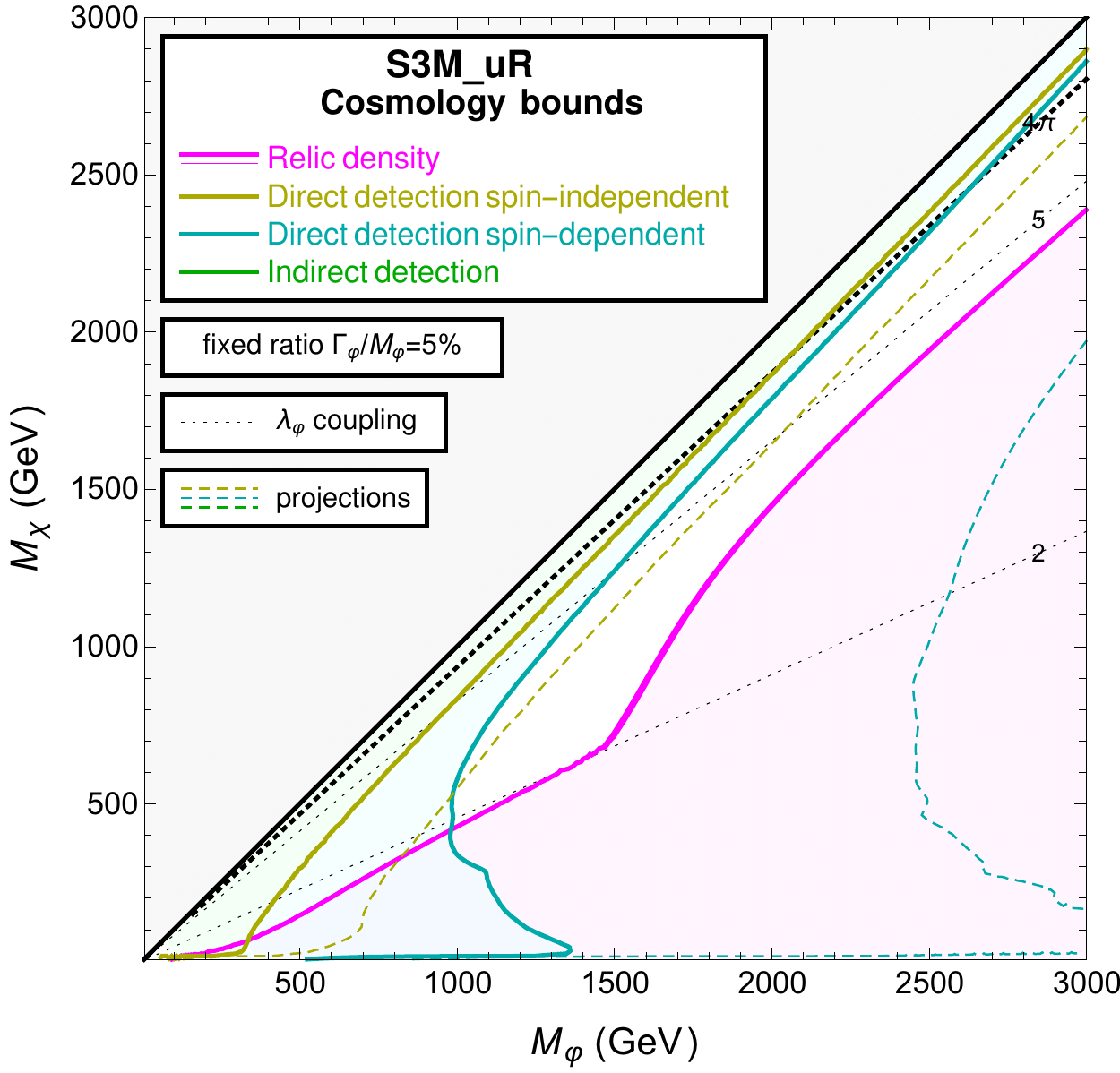}
\includegraphics[clip,width=.32\textwidth]{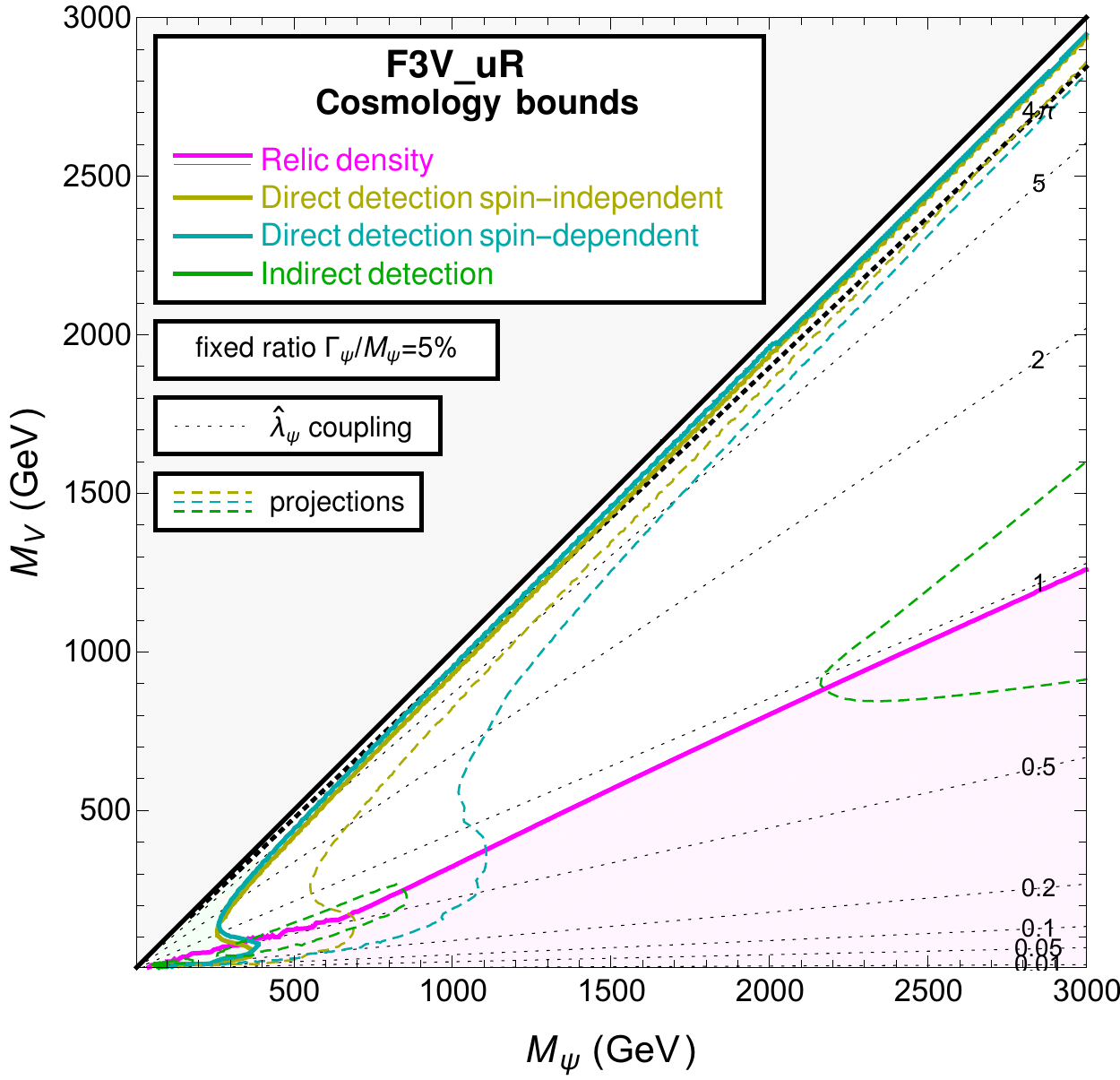}\\
\includegraphics[clip,width=.32\textwidth]{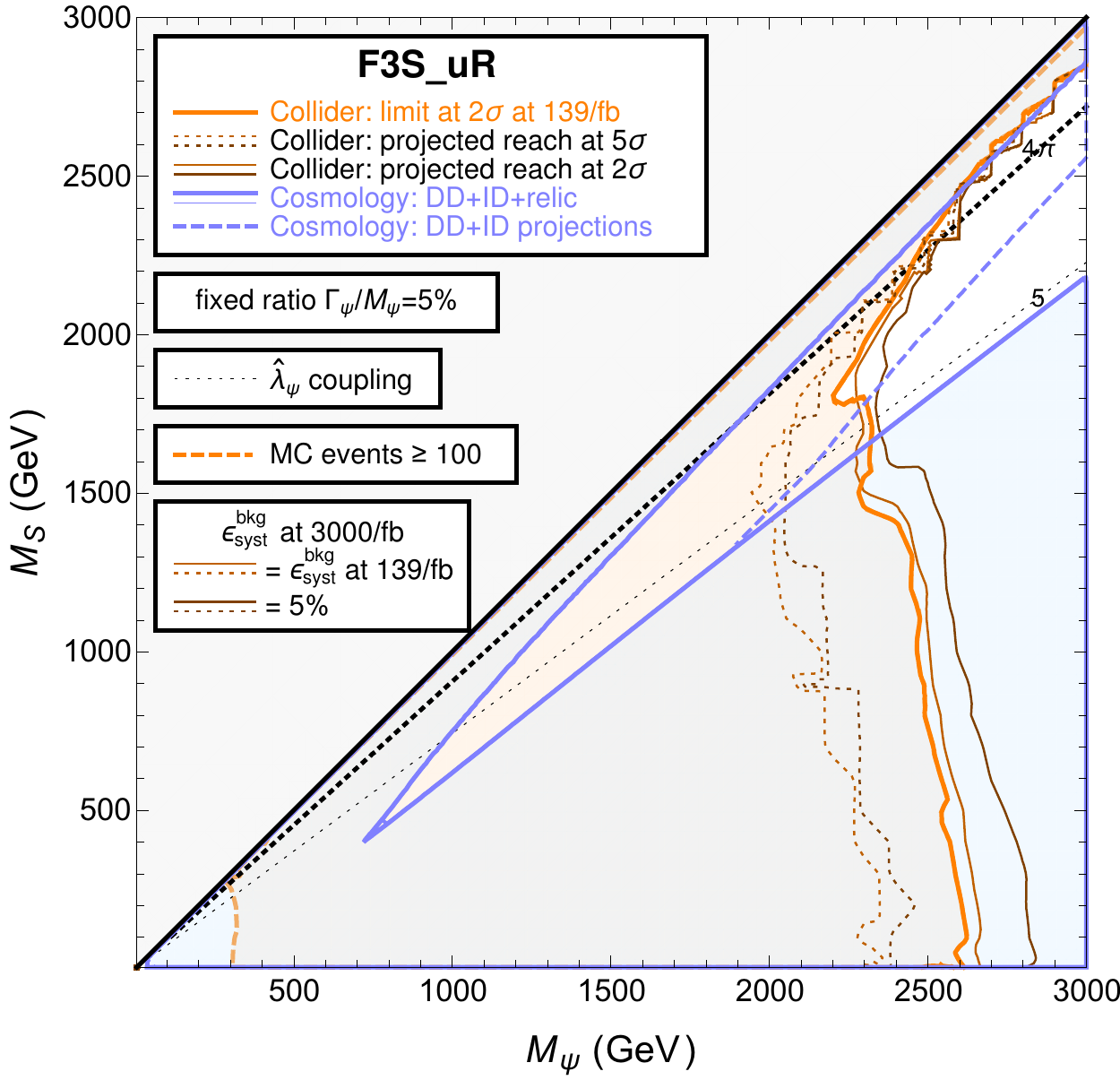}
\includegraphics[clip,width=.32\textwidth]{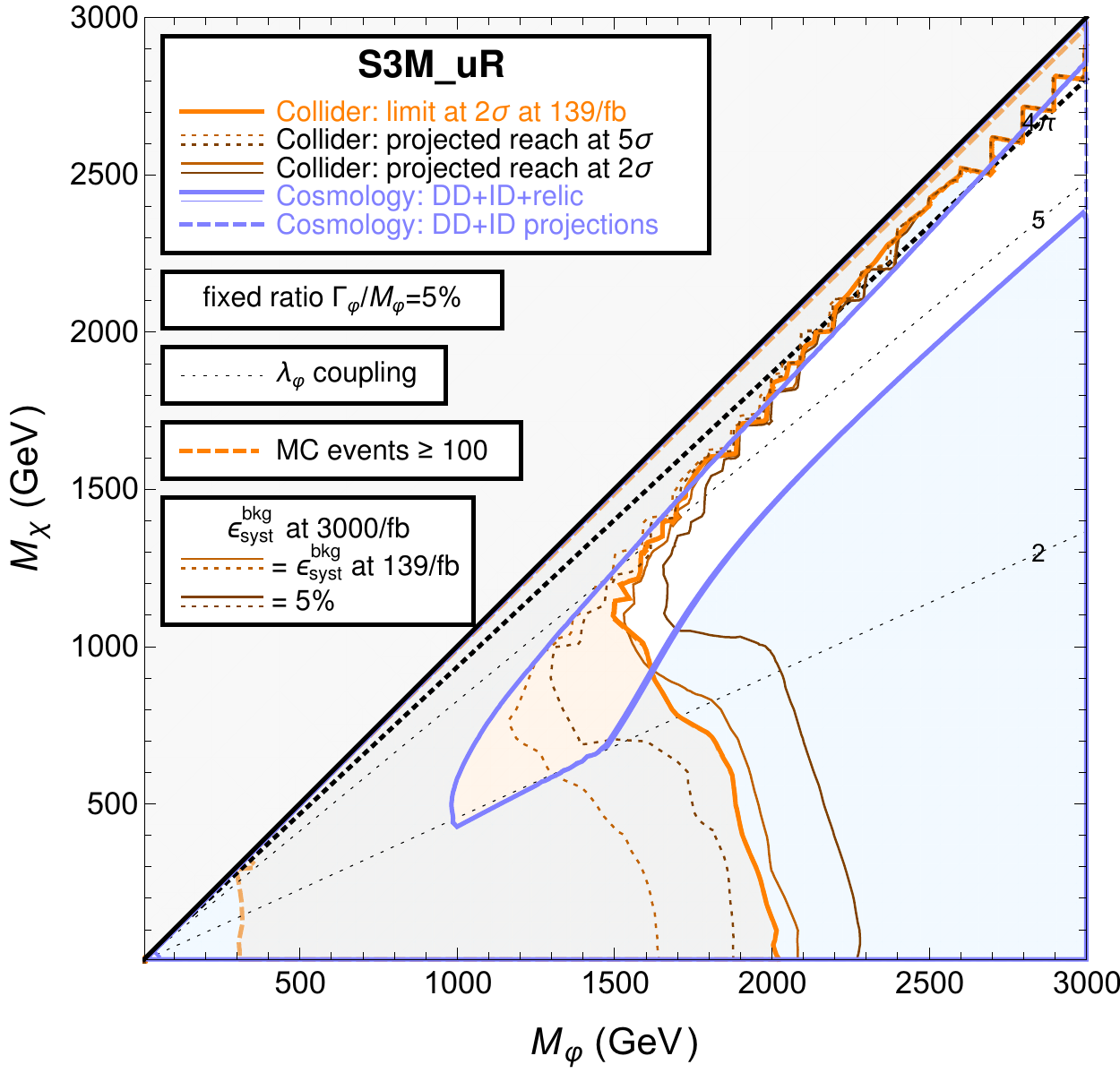}
\includegraphics[clip,width=.32\textwidth]{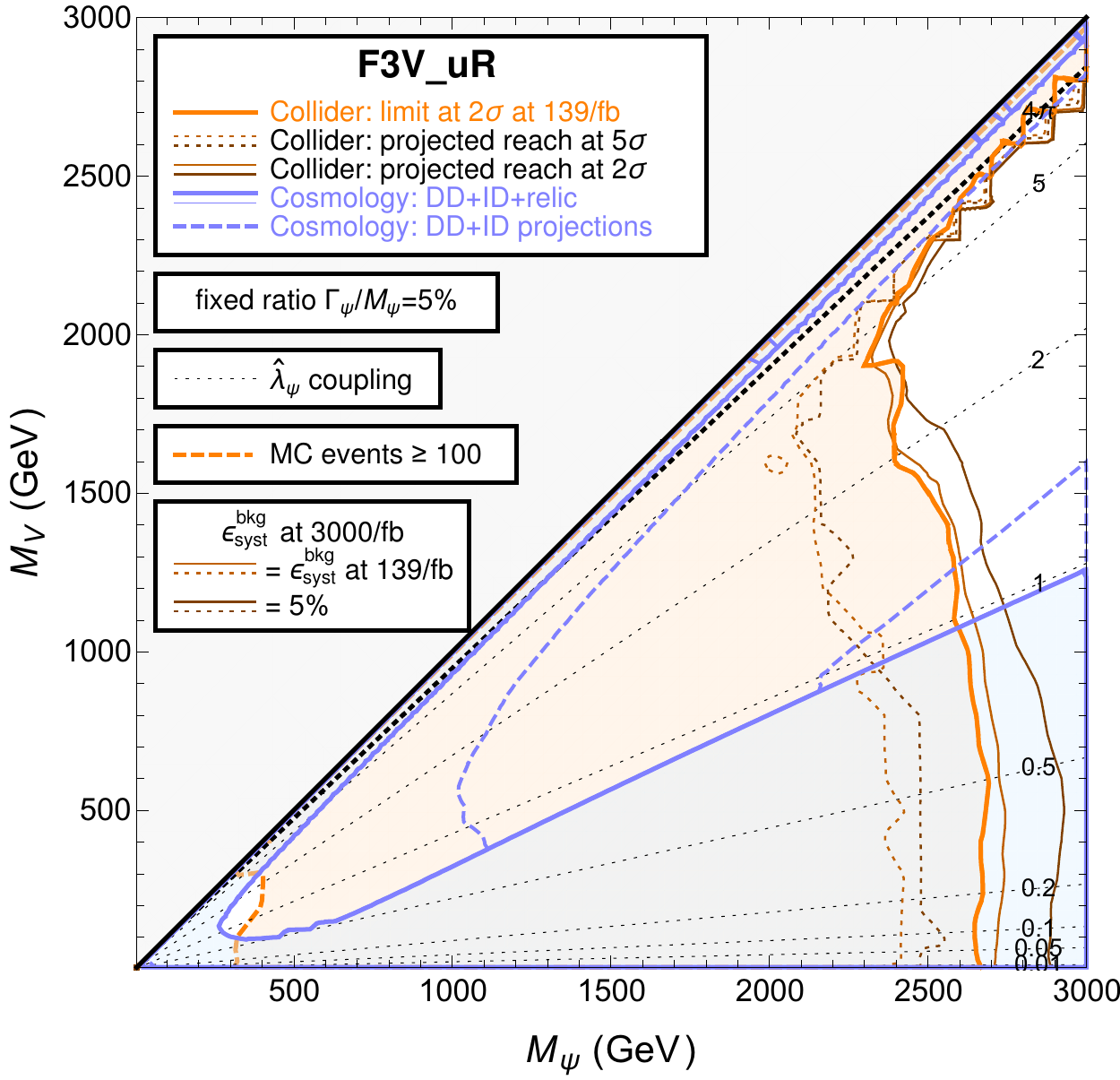}
\caption{
{\bfseries Top row:} Bounds arising from cosmological observations, represented
in the $(m_{X}, m_{Y})$ plane for a fixed mediator width over mass ratio of 5\%,
on the {\tt F3S\_uR} (left), {\tt S3M\_uR} (centre) and {\tt F3V\_uR} (right)
model parameter space. We allow for under-abundant dark matter, the band which
reproduces the relic density as measured by the Planck collaboration lying
between the (almost indistinguishable) thin and thick magenta lines. Future
projections of the constraints are provided as dashed lines. {\bfseries Bottom
row:} Combination of cosmological and collider bounds and their projections. The
collider projections correspond to exclusion and discovery reaches for a LHC
luminosity of 3000~fb$^{-1}$ ({\it i.e.} the HL-LHC phase) and assume a level of
systematics on the background either equal to the considered ATLAS search at
current luminosity or fixed to 5\%.}
\label{fig:all_combined}
\end{figure*}

We illustrate in \cref{fig:all_combined} the complementarity of the considered
cosmological and collider constraints on the models, after mapping the
cosmological bounds of \cref{sec:cosmo} onto an $(m_{X}, m_{Y})$ plane for a
fixed $\Gamma_Y/m_Y$ ratio of 5\%\footnote{Representing the results for a fixed
mediator width-over-mass ratio is only \textit{one} of the possibilities. We
could instead enforce a fixed $\lambda$ value. In this case, $\Gamma_Y/m_Y$
increases for large mediator and small dark matter masses. The NWA can thus
break down, depending on the value of the coupling and on the model, and
simulations relying on the factorisation of the mediator
production and decay processes, as traditionally performed, would lead to an
incorrect description of the signal kinematics.}.
However, contrary to the previous section, we allow for under-abundant dark
matter and therefore only consider the relic density constraint as an upper
bound. We hence implicitly assume the existence of some other dark matter
component, with different properties and interactions.

Under these assumptions, we obtain allowed parameter space regions for all
scenarios. These regions feature mediator masses greater than 1.5~TeV (scalar
mediator) and 2~TeV (fermion mediator), and a neither too compressed nor too
split new physics spectrum. The mediator mass is mostly constrained by
collider searches, while the dark matter mass is restricted by the combination
of the relic density (lower bound) and the interplay between the SD and SI
direct detection (upper bound) constraints. The only exception concerns the
small mass gap regime in which the collider constraints tend to be competitive
(despite of potentially non-perturbative couplings). For all scenarios, indirect
detection constraints are too weak to play any role. Gamma-ray fluxes are indeed
reduced by a $(\Omega h^2_{\rm model}/\Omega h^2_{\rm Planck})^2$ factor for
under-abundant dark matter, contrarily to the direct detection predictions that
are only linearly rescaled\footnote{Without such a rescaling (as when a
non-thermal mechanism is invoked to reproduce the dark matter
density~\cite{Allahverdi:2012wb}), the combination of current direct and
indirect detection bounds disfavours the parameter space regions which the
HL-LHC is sensitive to. We do not consider such a possibility in this work.}.

This collider-cosmology complementarity of constraints is compatible with the
nature of the considered experimental probes. Collider bounds are largely
dominated by the impact of the $YY$ channel. On the contrary, direct detection
experiments are more sensitive to scenarios featuring large couplings and/or a
small $X$/$Y$ mass splitting, as the direct detection cross section scales as
$\lambda^4$ and exhibits a polynomial in $r$ in its denominator. Moreover, when
allowing for under-abundant dark matter, the relic density favours large
couplings as well and opens the door to a much wider set of viable solutions.

In the same \cref{fig:all_combined}, we provide projections for future
experiments. We show projected $2\sigma$ exclusion and $5\sigma$ discovery
reaches for the HL-LHC, which corresponds to a luminosity of 3~ab$^{-1}$. We
extrapolate the current reach under two assumptions for the manner the
systematic uncertainties on the background $\epsilon_{\rm syst}^{\rm bgk}$ could
evolve. In a first case, we consider that it is the same as in the initially
considered 139~fb$^{-1}$ ATLAS search, while in the second case,
we assume that it reaches a floor of 5\% for each SR.
The results show that considering the recast cut-and-count ATLAS analysis, the
bounds will not improve significantly even with an optimistic assumption on the
systematics. Equivalently, this shows that the discovery reach is very close to
the current exclusion limits. Different, more complex, analysis strategies
should therefore be considered to better assess the potential of future searches
in probing a wider region of the parameter space. For example, the ATLAS search
that we have used in our analysis also includes a supersymmetry-inspired signal
region relying on a boosted decision tree, which we did not consider in our
model-independent approach.

On the other hand, the projected cosmological bounds have a much larger
potential. We present the expected sensitivity of future SI
(LZ~\cite{Mount:2017qzi}) and SD (LZ, PICO-500~\cite{Amole:2016pye} and
COSINUS~\cite{Angloher:2016ooq}) direct detection experiments, the latter being
extracted from ref.~\cite{Kang:2018odb} for the $\mathcal{O}_4$ operator (\ie\
for standard SD interactions). The interplay of the projected SD bounds and the
relic density constraints can completely exclude the {\tt S3M\_uR} scenario,
while the improvement of the SI bounds would drastically limit the options
allowed by the HL-LHC expectation for the {\tt F3S\_uR} model.
Similarly, high-energy gamma-ray experiments, such as CTA~\cite{Lefranc:2016fgn,
Acharyya:2020sbj} and SWGO~\cite{Viana:2019ucn}, and the LSST+Fermi-LAT
dSphs survey~\cite{Drlica-Wagner:2019xan} will be able to explore the model
parameter space well above the TeV regime, in a region that is out of reach of
LHC searches. In particular, projections for indirect detection has the largest
impact on the {\tt F3V\_uR} scenario, being the dominant constraint for large
mediator masses. It however still leaves a large window testable at the HL-LHC.

\section{Conclusions}\label{sec:conclusions}
We have performed a comprehensive analysis of cosmological and collider bounds
for three sets of $t$-channel simplified dark matter models in which the dark
matter is a real field. We have investigated the complementarity between the
different types of bounds and made projections for future collider and
cosmological experiments. Our findings show that most parameter spaces are
already strongly constrained by current bounds, and that future dark matter
direct and indirect detection probe have a large potential to cover the
still allowed regions of the parameter space. In this way, conclusive statements
on the phenomenological viability of the considered class of $t$-channel models
will be in order in the next decades.

One should however keep in mind that the models considered in this analysis are
simplified and model-independent constructions. While being representative of
different the\-o\-re\-ti\-cal\-ly-motivated
new physics scenarios, they necessarily lack non-minimal features, such as the
presence of more mediators, a multi-component dark matter spectrum, or a wider
range of interactions between the new particles and the SM. Such features can
change the picture by introducing, for example, interference contributions which
can weaken the constraints or effects due to large mediator widths which modify the final-state kinematics at colliders.

Finally, we did not investigate freeze-in dark matter scenarios, which we leave
for a separate work. This scenario is viable for tiny $\lambda$ values
of the order of $10^{-6}$ or smaller and might open up additional windows, as
for instance related to LLP searches at the LHC.

\section*{Acknowledgments}
We acknowledge J. Heisig, M. Kraemer and K. Mawatari for stimulating discussions during this study, and L. Lopez Honorez and M. Tytgat for their help in the validation procedure. LP work is supported by the Knut and Alice Wallenberg foundation under the SHIFT project, grant KAW 2017.0100. LP acknowledges the use of the IRIDIS 4 HPC Facility at the University of Southampton. CA is supported by the Innoviris  ATTRACT 2018 104 BECAP 2 agreement. HM is supported by the German Research Foundation DFG through the RTG 2497 and the CRC/Transregio 257. LM is supported by funding from the European Union's Horizon 2020 research and innovation programme as part of the Marie Sklodowska-Curie Innovative Training Network MCnetITN3 (grant agreement no. 722104).

\bibliographystyle{JHEP}
\bibliography{biblio}

\providecommand{\href}[2]{#2}\begingroup\raggedright\begin{thebibliography}{10}

\bibitem{Bertone:2010zza}
J.~Silk et~al., \emph{{Particle Dark Matter: Observations, Models and
  Searches}}. Cambridge Univ. Press, Cambridge, 2010,
  \href{https://doi.org/10.1017/CBO9780511770739}{10.1017/CBO9780511770739}.

\bibitem{Fox:2012ru}
P.~J. Fox and C.~Williams, \emph{{Next-to-Leading Order Predictions for Dark
  Matter Production at Hadron Colliders}},
  \href{https://doi.org/10.1103/PhysRevD.87.054030}{\emph{Phys. Rev.}
  {\bfseries D87} (2013) 054030}
  [\href{https://arxiv.org/abs/1211.6390}{{\ttfamily 1211.6390}}].

\bibitem{Haisch:2013ata}
U.~Haisch, F.~Kahlhoefer and E.~Re, \emph{{QCD effects in mono-jet searches for
  dark matter}}, \href{https://doi.org/10.1007/JHEP12(2013)007}{\emph{JHEP}
  {\bfseries 12} (2013) 007} [\href{https://arxiv.org/abs/1310.4491}{{\ttfamily
  1310.4491}}].

\bibitem{Backovic:2015soa}
M.~Backovic, M.~Kraemer, F.~Maltoni, A.~Martini, K.~Mawatari and M.~Pellen,
  \emph{{Higher-order QCD predictions for dark matter production at the LHC in
  simplified models with s-channel mediators}},
  \href{https://doi.org/10.1140/epjc/s10052-015-3700-6}{\emph{Eur. Phys. J.}
  {\bfseries C75} (2015) 482}
  [\href{https://arxiv.org/abs/1508.05327}{{\ttfamily 1508.05327}}].

\bibitem{Arina:2020udz}
C.~Arina, B.~Fuks and L.~Mantani, \emph{{A universal framework for t-channel
  dark matter models}},
  \href{https://doi.org/10.1140/epjc/s10052-020-7933-7}{\emph{Eur. Phys. J. C}
  {\bfseries 80} (2020) 409}
  [\href{https://arxiv.org/abs/2001.05024}{{\ttfamily 2001.05024}}].

\bibitem{Degrande:2011ua}
C.~Degrande, C.~Duhr, B.~Fuks, D.~Grellscheid, O.~Mattelaer and T.~Reiter,
  \emph{{UFO - The Universal FeynRules Output}},
  \href{https://doi.org/10.1016/j.cpc.2012.01.022}{\emph{Comput. Phys. Commun.}
  {\bfseries 183} (2012) 1201}
  [\href{https://arxiv.org/abs/1108.2040}{{\ttfamily 1108.2040}}].

\bibitem{Alwall:2014hca}
J.~Alwall, R.~Frederix, S.~Frixione, V.~Hirschi, F.~Maltoni, O.~Mattelaer
  et~al., \emph{{The automated computation of tree-level and next-to-leading
  order differential cross sections, and their matching to parton shower
  simulations}}, \href{https://doi.org/10.1007/JHEP07(2014)079}{\emph{JHEP}
  {\bfseries 07} (2014) 079} [\href{https://arxiv.org/abs/1405.0301}{{\ttfamily
  1405.0301}}].

\bibitem{Belyaev:2012qa}
A.~Belyaev, N.~D. Christensen and A.~Pukhov, \emph{{CalcHEP 3.4 for collider
  physics within and beyond the Standard Model}},
  \href{https://doi.org/10.1016/j.cpc.2013.01.014}{\emph{Comput. Phys. Commun.}
  {\bfseries 184} (2013) 1729}
  [\href{https://arxiv.org/abs/1207.6082}{{\ttfamily 1207.6082}}].

\bibitem{Ambrogi:2018jqj}
F.~Ambrogi et~al., \emph{{MadDM v.3.0: a Comprehensive Tool for Dark Matter
  Studies}}, \href{https://doi.org/10.1016/j.dark.2018.11.009}{\emph{Phys. Dark
  Univ.} {\bfseries 24} (2019) 100249}
  [\href{https://arxiv.org/abs/1804.00044}{{\ttfamily 1804.00044}}].

\bibitem{Belanger:2018mqt}
G.~Belanger, F.~Boudjema, A.~Goudelis, A.~Pukhov and B.~Zaldivar,
  \emph{{micrOMEGAs5.0 : Freeze-in}},
  \href{https://doi.org/10.1016/j.cpc.2018.04.027}{\emph{Comput. Phys. Commun.}
  {\bfseries 231} (2018) 173}
  [\href{https://arxiv.org/abs/1801.03509}{{\ttfamily 1801.03509}}].

\bibitem{Alloul:2013bka}
A.~Alloul, N.~D. Christensen, C.~Degrande, C.~Duhr and B.~Fuks,
  \emph{{FeynRules 2.0 - A complete toolbox for tree-level phenomenology}},
  \href{https://doi.org/10.1016/j.cpc.2014.04.012}{\emph{Comput. Phys. Commun.}
  {\bfseries 185} (2014) 2250}
  [\href{https://arxiv.org/abs/1310.1921}{{\ttfamily 1310.1921}}].

\bibitem{Berdine:2007uv}
D.~Berdine, N.~Kauer and D.~Rainwater, \emph{{Breakdown of the Narrow Width
  Approximation for New Physics}},
  \href{https://doi.org/10.1103/PhysRevLett.99.111601}{\emph{Phys. Rev. Lett.}
  {\bfseries 99} (2007) 111601}
  [\href{https://arxiv.org/abs/hep-ph/0703058}{{\ttfamily hep-ph/0703058}}].

\bibitem{Ball:2014uwa}
{\scshape NNPDF} collaboration, \emph{{Parton distributions for the LHC Run
  II}}, \href{https://doi.org/10.1007/JHEP04(2015)040}{\emph{JHEP} {\bfseries
  04} (2015) 040} [\href{https://arxiv.org/abs/1410.8849}{{\ttfamily
  1410.8849}}].

\bibitem{Buckley:2014ana}
A.~Buckley, J.~Ferrando, S.~Lloyd, K.~Nordstr\"om, B.~Page, M.~R\"ufenacht
  et~al., \emph{{LHAPDF6: parton density access in the LHC precision era}},
  \href{https://doi.org/10.1140/epjc/s10052-015-3318-8}{\emph{Eur. Phys. J. C}
  {\bfseries 75} (2015) 132} [\href{https://arxiv.org/abs/1412.7420}{{\ttfamily
  1412.7420}}].

\bibitem{ATLAS:2019vcq}
{\scshape ATLAS} collaboration, \emph{{Search for squarks and gluinos in final
  states with jets and missing transverse momentum using 139 fb$^{-1}$ of
  $\sqrt{s}$ =13 TeV $pp$ collision data with the ATLAS detector}},
  ATLAS-CONF-2019-040.

\bibitem{Conte:2018vmg}
E.~Conte and B.~Fuks, \emph{{Confronting new physics theories to LHC data with
  MADANALYSIS 5}}, \href{https://doi.org/10.1142/S0217751X18300272}{\emph{Int.
  J. Mod. Phys.} {\bfseries A33} (2018) 1830027}
  [\href{https://arxiv.org/abs/1808.00480}{{\ttfamily 1808.00480}}].

\bibitem{ATLASCONF2019040recast}
F.~Ambrogi, \emph{{MadAnalysis 5 recast of ATLAS-CONF-2019-040}},  2019.
\newblock DOI: 10.7484/INSPIREHEP.DATA.45EF.23SB.

\bibitem{Colucci:2018vxz}
S.~Colucci, B.~Fuks, F.~Giacchino, L.~Lopez~Honorez, M.~H.~G. Tytgat and
  J.~Vandecasteele, \emph{{Top-philic Vector-Like Portal to Scalar Dark
  Matter}}, \href{https://doi.org/10.1103/PhysRevD.98.035002}{\emph{Phys. Rev.}
  {\bfseries D98} (2018) 035002}
  [\href{https://arxiv.org/abs/1804.05068}{{\ttfamily 1804.05068}}].

\bibitem{Garny:2018icg}
M.~Garny, J.~Heisig, M.~Hufnagel and B.~Luelf, \emph{{Top-philic dark matter
  within and beyond the WIMP paradigm}},
  \href{https://doi.org/10.1103/PhysRevD.97.075002}{\emph{Phys. Rev.}
  {\bfseries D97} (2018) 075002}
  [\href{https://arxiv.org/abs/1802.00814}{{\ttfamily 1802.00814}}].

\bibitem{Read:2002hq}
A.~L. Read, \emph{{Presentation of search results: The CL(s) technique}},
  \href{https://doi.org/10.1088/0954-3899/28/10/313}{\emph{J. Phys.} {\bfseries
  G28} (2002) 2693}.

\bibitem{Araz:2019otb}
J.~Y. Araz, M.~Frank and B.~Fuks, \emph{{Reinterpreting the results of the LHC
  with MadAnalysis 5: uncertainties and higher-luminosity estimates}},
  \href{https://doi.org/10.1140/epjc/s10052-020-8076-6}{\emph{Eur. Phys. J. C}
  {\bfseries 80} (2020) 531}
  [\href{https://arxiv.org/abs/1910.11418}{{\ttfamily 1910.11418}}].

\bibitem{Aghanim:2018eyx}
{\scshape Planck} collaboration, \emph{{Planck 2018 results. VI. Cosmological
  parameters}},
  \href{https://doi.org/10.1051/0004-6361/201833910}{\emph{Astron. Astrophys.}
  {\bfseries 641} (2020) A6}
  [\href{https://arxiv.org/abs/1807.06209}{{\ttfamily 1807.06209}}].

\bibitem{Aprile:2018dbl}
{\scshape XENON} collaboration, \emph{{Dark Matter Search Results from a One
  Ton-Year Exposure of XENON1T}},
  \href{https://doi.org/10.1103/PhysRevLett.121.111302}{\emph{Phys. Rev. Lett.}
  {\bfseries 121} (2018) 111302}
  [\href{https://arxiv.org/abs/1805.12562}{{\ttfamily 1805.12562}}].

\bibitem{Ackermann:2015lka}
{\scshape Fermi-LAT} collaboration, \emph{{Updated search for spectral lines
  from Galactic dark matter interactions with pass 8 data from the Fermi Large
  Area Telescope}},
  \href{https://doi.org/10.1103/PhysRevD.91.122002}{\emph{Phys. Rev.}
  {\bfseries D91} (2015) 122002}
  [\href{https://arxiv.org/abs/1506.00013}{{\ttfamily 1506.00013}}].

\bibitem{Abdallah:2018qtu}
{\scshape HESS} collaboration, \emph{{Search for $\gamma$-Ray Line Signals from
  Dark Matter Annihilations in the Inner Galactic Halo from 10 Years of
  Observations with H.E.S.S.}},
  \href{https://doi.org/10.1103/PhysRevLett.120.201101}{\emph{Phys. Rev. Lett.}
  {\bfseries 120} (2018) 201101}
  [\href{https://arxiv.org/abs/1805.05741}{{\ttfamily 1805.05741}}].

\bibitem{Amole:2017dex}
{\scshape PICO} collaboration, \emph{{Dark Matter Search Results from the
  PICO-60 C$_3$F$_8$ Bubble Chamber}},
  \href{https://doi.org/10.1103/PhysRevLett.118.251301}{\emph{Phys. Rev. Lett.}
  {\bfseries 118} (2017) 251301}
  [\href{https://arxiv.org/abs/1702.07666}{{\ttfamily 1702.07666}}].

\bibitem{Fermi-LAT:2016uux}
{\scshape DES, Fermi-LAT} collaboration, \emph{{Searching for Dark Matter
  Annihilation in Recently Discovered Milky Way Satellites with Fermi-LAT}},
  \href{https://doi.org/10.3847/1538-4357/834/2/110}{\emph{Astrophys. J.}
  {\bfseries 834} (2017) 110}
  [\href{https://arxiv.org/abs/1611.03184}{{\ttfamily 1611.03184}}].

\bibitem{Toma:2013bka}
T.~Toma, \emph{{Internal Bremsstrahlung Signature of Real Scalar Dark Matter
  and Consistency with Thermal Relic Density}},
  \href{https://doi.org/10.1103/PhysRevLett.111.091301}{\emph{Phys. Rev. Lett.}
  {\bfseries 111} (2013) 091301}
  [\href{https://arxiv.org/abs/1307.6181}{{\ttfamily 1307.6181}}].

\bibitem{Giacchino:2013bta}
F.~Giacchino, L.~Lopez-Honorez and M.~H.~G. Tytgat, \emph{{Scalar Dark Matter
  Models with Significant Internal Bremsstrahlung}},
  \href{https://doi.org/10.1088/1475-7516/2013/10/025}{\emph{JCAP} {\bfseries
  1310} (2013) 025} [\href{https://arxiv.org/abs/1307.6480}{{\ttfamily
  1307.6480}}].

\bibitem{Giacchino:2014moa}
F.~Giacchino, L.~Lopez-Honorez and M.~H.~G. Tytgat, \emph{{Bremsstrahlung and
  Gamma Ray Lines in 3 Scenarios of Dark Matter Annihilation}},
  \href{https://doi.org/10.1088/1475-7516/2014/08/046}{\emph{JCAP} {\bfseries
  1408} (2014) 046} [\href{https://arxiv.org/abs/1405.6921}{{\ttfamily
  1405.6921}}].

\bibitem{Giacchino:2015hvk}
F.~Giacchino, A.~Ibarra, L.~Lopez~Honorez, M.~H.~G. Tytgat and S.~Wild,
  \emph{{Signatures from Scalar Dark Matter with a Vector-like Quark
  Mediator}}, \href{https://doi.org/10.1088/1475-7516/2016/02/002}{\emph{JCAP}
  {\bfseries 1602} (2016) 002}
  [\href{https://arxiv.org/abs/1511.04452}{{\ttfamily 1511.04452}}].

\bibitem{Biondini:2019int}
S.~Biondini and S.~Vogl, \emph{{Scalar dark matter coannihilating with a
  coloured fermion}},
  \href{https://doi.org/10.1007/JHEP11(2019)147}{\emph{JHEP} {\bfseries 11}
  (2019) 147} [\href{https://arxiv.org/abs/1907.05766}{{\ttfamily
  1907.05766}}].

\bibitem{Biondini:2018ovz}
S.~Biondini and S.~Vogl, \emph{{Coloured coannihilations: Dark matter
  phenomenology meets non-relativistic EFTs}},
  \href{https://doi.org/10.1007/JHEP02(2019)016}{\emph{JHEP} {\bfseries 02}
  (2019) 016} [\href{https://arxiv.org/abs/1811.02581}{{\ttfamily
  1811.02581}}].

\bibitem{Ibarra:2014qma}
A.~Ibarra, T.~Toma, M.~Totzauer and S.~Wild, \emph{{Sharp Gamma-ray Spectral
  Features from Scalar Dark Matter Annihilations}},
  \href{https://doi.org/10.1103/PhysRevD.90.043526}{\emph{Phys. Rev.}
  {\bfseries D90} (2014) 043526}
  [\href{https://arxiv.org/abs/1405.6917}{{\ttfamily 1405.6917}}].

\bibitem{Hisano:2015bma}
J.~Hisano, R.~Nagai and N.~Nagata, \emph{{Effective Theories for Dark Matter
  Nucleon Scattering}},
  \href{https://doi.org/10.1007/JHEP05(2015)037}{\emph{JHEP} {\bfseries 05}
  (2015) 037} [\href{https://arxiv.org/abs/1502.02244}{{\ttfamily
  1502.02244}}].

\bibitem{Mohan:2019zrk}
K.~A. Mohan, D.~Sengupta, T.~M.~P. Tait, B.~Yan and C.~P. Yuan, \emph{{Direct
  Detection and LHC constraints on a $t$-Channel Simplified Model of Majorana
  Dark Matter at One Loop}},
  \href{https://doi.org/10.1007/JHEP05(2019)115}{\emph{JHEP} {\bfseries 05}
  (2019) 115} [\href{https://arxiv.org/abs/1903.05650}{{\ttfamily
  1903.05650}}].

\bibitem{Einasto:2009zd}
J.~Einasto, \emph{{Dark Matter}},  in \emph{{Astronomy and Astrophysics 2010,
  [Eds. Oddbjorn Engvold, Rolf Stabell, Bozena Czerny, John Lattanzio], in
  Encyclopedia of Life Support Systems (EOLSS), Developed under the Auspices of
  the UNESCO, Eolss Publishers, Oxford ,UK}}, 2009,
  \href{https://arxiv.org/abs/0901.0632}{{\ttfamily 0901.0632}}.

\bibitem{Garny:2013ama}
M.~Garny, A.~Ibarra, M.~Pato and S.~Vogl, \emph{{Internal bremsstrahlung
  signatures in light of direct dark matter searches}},
  \href{https://doi.org/10.1088/1475-7516/2013/12/046}{\emph{JCAP} {\bfseries
  1312} (2013) 046} [\href{https://arxiv.org/abs/1306.6342}{{\ttfamily
  1306.6342}}].

\bibitem{Alimena:2019zri}
J.~Alimena et~al., \emph{{Searching for Long-Lived Particles beyond the
  Standard Model at the Large Hadron Collider}},
  \href{https://doi.org/10.1088/1361-6471/ab4574}{\emph{J. Phys. G} {\bfseries
  47} (2020) 090501} [\href{https://arxiv.org/abs/1903.04497}{{\ttfamily
  1903.04497}}].

\bibitem{Allahverdi:2012wb}
R.~Allahverdi, B.~Dutta and K.~Sinha, \emph{{Non-thermal Higgsino Dark Matter:
  Cosmological Motivations and Implications for a 125 GeV Higgs}},
  \href{https://doi.org/10.1103/PhysRevD.86.095016}{\emph{Phys. Rev. D}
  {\bfseries 86} (2012) 095016}
  [\href{https://arxiv.org/abs/1208.0115}{{\ttfamily 1208.0115}}].

\bibitem{Mount:2017qzi}
B.~Mount et~al., \emph{{LUX-ZEPLIN (LZ) Technical Design Report}},
  \href{https://arxiv.org/abs/1703.09144}{{\ttfamily 1703.09144}}.

\bibitem{Amole:2016pye}
{\scshape PICO} collaboration, \emph{{Improved dark matter search results from
  PICO-2L Run 2}},
  \href{https://doi.org/10.1103/PhysRevD.93.061101}{\emph{Phys. Rev. D}
  {\bfseries 93} (2016) 061101}
  [\href{https://arxiv.org/abs/1601.03729}{{\ttfamily 1601.03729}}].

\bibitem{Angloher:2016ooq}
G.~Angloher et~al., \emph{{The COSINUS project - perspectives of a NaI
  scintillating calorimeter for dark matter search}},
  \href{https://doi.org/10.1140/epjc/s10052-016-4278-3}{\emph{Eur. Phys. J. C}
  {\bfseries 76} (2016) 441}
  [\href{https://arxiv.org/abs/1603.02214}{{\ttfamily 1603.02214}}].

\bibitem{Kang:2018odb}
S.~Kang, S.~Scopel, G.~Tomar and J.-H. Yoon, \emph{{Present and projected
  sensitivities of Dark Matter direct detection experiments to effective
  WIMP-nucleus couplings}},
  \href{https://doi.org/10.1016/j.astropartphys.2019.02.006}{\emph{Astropart.
  Phys.} {\bfseries 109} (2019) 50}
  [\href{https://arxiv.org/abs/1805.06113}{{\ttfamily 1805.06113}}].

\bibitem{Lefranc:2016fgn}
V.~Lefranc, E.~Moulin, P.~Panci, F.~Sala and J.~Silk, \emph{{Dark Matter in
  $\gamma$ lines: Galactic Center vs dwarf galaxies}},
  \href{https://doi.org/10.1088/1475-7516/2016/09/043}{\emph{JCAP} {\bfseries
  09} (2016) 043} [\href{https://arxiv.org/abs/1608.00786}{{\ttfamily
  1608.00786}}].

\bibitem{Acharyya:2020sbj}
{\scshape CTA} collaboration, \emph{{Pre-construction estimates of the
  Cherenkov Telescope Array sensitivity to a dark matter signal from the
  Galactic centre}},  \href{https://arxiv.org/abs/2007.16129}{{\ttfamily
  2007.16129}}.

\bibitem{Viana:2019ucn}
A.~Viana, H.~Schoorlemmer, A.~Albert, V.~de~Souza, J.~P. Harding and J.~Hinton,
  \emph{{Searching for Dark Matter in the Galactic Halo with a Wide Field of
  View TeV Gamma-ray Observatory in the Southern Hemisphere}},
  \href{https://doi.org/10.1088/1475-7516/2019/12/061}{\emph{JCAP} {\bfseries
  12} (2019) 061} [\href{https://arxiv.org/abs/1906.03353}{{\ttfamily
  1906.03353}}].

\bibitem{Drlica-Wagner:2019xan}
{\scshape LSST Dark Matter Group} collaboration, \emph{{Probing the Fundamental
  Nature of Dark Matter with the Large Synoptic Survey Telescope}},
  \href{https://arxiv.org/abs/1902.01055}{{\ttfamily 1902.01055}}.

\end{thebibliography}\endgroup

\end{document}